\documentclass[journal]{IEEEtran}

\ifCLASSINFOpdf
\usepackage[pdftex]{graphicx}
\else
\fi

\usepackage[cmex10]{amsmath}

\usepackage{graphicx}
\usepackage{subfigure}
\usepackage{algorithmic}

\usepackage[utf8]{inputenc}
\usepackage[english]{babel}
\usepackage{amsthm}
\usepackage{stfloats}

\usepackage{threeparttable}
\usepackage{newclude}
\usepackage{multirow}
\usepackage{amsfonts}
\usepackage{amsthm}
\usepackage{verbatim}
\usepackage{xcolor}
\usepackage{textcomp}		% for the "cents" symbol
\usepackage{epstopdf}
\usepackage{textcomp}
\usepackage[caption=false]{subfig}
\usepackage{tabularx}
\newcolumntype{P}[1]{>{\centering\arraybackslash}p{#1}}
\newcolumntype{M}[1]{>{\centering\arraybackslash}m{#1}}
\newtheorem{theorem}{\textbf{Theorem}}
\newtheorem{assumption}{\textbf{Assumption}}
\newtheorem{lemma}{\textbf{Lemma}}

\usepackage{color,soul}

\usepackage{cite}  %put it in the last sentence to make the continuous ref []-[]
\usepackage{algorithm} %format of the algorithm
\usepackage{algorithmic} %format of the algorithm
\begin{document}
	%
	% paper title
	% can use linebreaks \\ within to get better formatting as desired
	% Do not put math or special symbols in the title.
	%\title{Reducing Non-detection Zone of Islanding Detection Schemes in Hybrid Distributed Generation}

	\title{Robust Hidden Topology Identification in Distribution Systems}
	
	% Based on Hardware-in-the-loop and Intelligent Relays}
	%
	%
	% author names and IEEE memberships
	% note positions of commas and nonbreaking spaces ( ~ ) LaTeX will not break
	% a structure at a ~ so this keeps an author's name from being broken across
	% two lines.
	% use \thanks{} to gain access to the first footnote area
	% a separate \thanks must be used for each paragraph as LaTeX2e's \thanks
	% was not built to handle multiple paragraphs
	%

\author{
Haoran~Li,~\IEEEmembership{Student Member,~IEEE,}
Yang~Weng,~\IEEEmembership{Member,~IEEE,}
Yizheng~Liao,~\IEEEmembership{Student Member,~IEEE,}
Brian~Keel,~\IEEEmembership{Senior Member,~IEEE,}
Kenneth E. Brown,~\IEEEmembership{Member,~IEEE}
 \vspace{-5mm}
\thanks{Haoran Li and Yang Weng are with the Department of Electrical, Computer and Energy Engineering, Arizona State University, Tempe, AZ, 85281 USA e-mail: \mbox{\{lhaoran,yang.weng\}@asu.edu};~
Y. Liao is with Department of Civil and Environmental Engineering, Stanford University, Stanford, CA, 94305 USA e-mail: yzliao@stanford.edu;~Brian~Keel and Kenneth E. Brown are with Salt River Project, Phoenix, AZ, 77005 USA e-mail: \mbox{\{brian.keel,Ken.Brown\}@srpnet.com}.}
\vspace{-5mm}}		

		%\thanks{The authors are with the Department
		%of Electrical and Computer Engineering, McGill University, Montreal, QC, H3A 0E9 Canada (e-mail: qiushi.cui@mail.mcgill.ca;khalil.elarroudi@mail.mcgill.ca;geza.joos@mcgill.ca)}.

		%qiushi.cui@mail.mcgill.ca,
		%\IEEEauthorrefmark{2}khalil.elarroudi@mail.mcgill.ca,
		%\IEEEauthorrefmark{3}geza.joos@mcgill.ca}}

%		\thanks{Manuscript received February 22, 2016; revised August 17, 2016.}

	% use for special paper notices
	%\IEEEspecialpapernotice{(Invited Paper)}

% make the title area
\maketitle

	% As a general rule, do not put math, special symbols or citations
	% in the abstract or keywords.
\begin{abstract}

% With more distributed energy resources (DERs) connected to distribution grids, new generation and load behaviors appear in the system. For further studies like stability analysis and demand response, identifying topologies is the basic step. However, operators usually cannot know a complete topology because of outdated connectivity information. Luckily, grid-structure information is reflected in voltage and current measurements of smart sensors. While these sensors are often limited to terminal nodes due to high cost, the line impedance between every two terminal nodes represents a unique path and these paths jointly determine the tree structure. This paper first proposes the line impedance estimation with terminal measurements. Based on these impedances, a recursive structure-learning algorithm is illustrated to identify the topology. In impedance estimation, the statistical correlations of current for different buses lead to high estimation errors. To eliminate correlations, this paper proves Cholesky whitening to be a suitable way in small grids. As for large systems, the paper applies a network reduction technique to decrease the size and guarantee the accuracy. Numerical performance is verified in multiple-size distribution grids with both simulation and real-world data.
With more distributed energy resources (DERs) connected to distribution grids, better monitoring and control are needed, where %like stability analysis and demand response,
identifying the topology accurately is the prerequisite. However, due to frequent re-configurations, operators usually cannot know a complete structure in distribution grids. Luckily, the growing data from smart sensors, restricted by Ohm's law, provides the possibility of topology inference. In this paper, we show how line parameters of Ohm's equation can be estimated for topology identification even when there are hidden nodes. %.
%\textcolor{red}{What is the relationship between the last sentence and the next sentence?} Nonetheless, sensors are often limited to end-user nodes due to high cost.
%Based on it, this paper shows that even partial line impedance from end user measurements are enough to find a tree topology.
%Specifically, beginning with end-user nodes, we aim at introducing a robust learning algorithm to
Specifically, the introduced learning method recursively conducts hidden-node detection and impedance calculation. %For the algorithm input, a line-impedance estimation among end nodes is employed. In this estimation,
However, the assumptions on uncorrelated data, availability of phasor measurements, and a balanced system, are not met in practices, causing large errors.
%\textcolor{red}{Why you find error, but would like to remove correlation?}
To resolve these problems, we employ Cholesky whitening first with a proof for measurement decorrelations. For increasing robustness further, we show how to handle practical scenarios when only measurement magnitudes are available or when the grid is three-phase unbalanced. Numerical performance is verified on multi-size distribution grids with both simulation and real-world data.
\end{abstract}

	% Note that keywords are not normally used for peerreview papers.

% \begin{IEEEkeywords}
% Distributed energy resources, distribution grids, hidden nodes, Cholesky whitening, network reduction, structure learning, line-impedance estimation
% \end{IEEEkeywords}

\IEEEpeerreviewmaketitle
\vspace{-8mm}	
\section{Introduction}
% * <liaoyizheng@gmail.com> 2019-01-02T06:44:57.116Z:
%
% > Introduction}
% I think the paper needs to clear out the logics among each section. Currently, it is difficult to follow. Also, there are too many mathmatical notations in this paper. It is better to have some consistent rules to write them or give a list of all notations as a reference.
%
% ^.
\label{section1}		
\IEEEPARstart{D}istributed energy resources (DERs) are broadly defined as renewable energy sources, electricity storage, and intelligent loads. They can offer more controllability for system operators and more choices for end-users. Furthermore, proper deployment of DERs brings economic benefits, such as reduction of network investment and increase clean energy share \cite{ref:dnv2014review}. Therefore, DER penetration has a consistent increase.
The New York State Energy Research \& Development Authority (NYSERDA) estimates a total $10,745$ GWh of commercial PV by $2030$ for the U.S. \cite{ref:dnv2014review}. However, numerous challenges also come. For example, DERs like rooftop solar panels can generate inverse power \cite{ref:weng2017d}. High penetration of PVs affects instant system power balancing \cite{ref:Lew2017t}. For the low-voltage network, DERs can cause the voltage rise and threaten the network reliability \cite{ref:Ferreira2013D}. Thus, power engineers need new monitoring, control and operating methods to face these profound changes.
% * <liaoyizheng@gmail.com> 2019-01-02T06:05:54.046Z:
%
% > \IEEEPARstart{D} do not need to use this env
%
%
% ^.

Distribution grid topology is a foundation for intelligent control and operation such as power flow study\cite{ref:Cui2014F} and stability analysis. However, the network structures aren't always available. Firstly, distribution grid reconfigures frequently. For example, in a city distribution network, routine reconfiguration helps to achieve a good radial topology from a meshed network \cite{ref:Rudin2012m, ref:Rudin2014a}. The frequency of a topology change ranges from once per eight hours with PVs\cite{ref:Jabr2014m} or once a month for medium-voltage grids \cite{ref:Fajardo2008r}  to once a season \cite{ref:Bueno2004d}. Secondly, some special changes, like the outages and manual maintenance, may not be reported immediately \cite{ref:weng2017d}. Further, considering the high cost, instruments like topology sensors aren't widely installed in the system. Finally, complete information about new DER components may be unavailable. For example, some plug-and-play components belong to users and the utility doesn't have the right to reach the breaker or switch information \cite{ref:weng2017d}.

Fortunately, there are growing real-time measurements in distribution systems. For example, metering devices like micro-PMUs \cite{ref:Meier2014d}, frequency measurement devices \cite{ref:Zhong2005p}, advanced sensors, and advanced metering infrastructure (AMI) are continuously being applied to distribution grids \cite{ref:Weng2015p, ref:Yu2015p}. Those meters are largely employed to enhance the observability of distribution systems \cite{ref:Bhela2018E}. Especially, topology detection is implemented via measurements, including voltage, current\cite{ref:Meier2017p} and load data.

For this reason, structure learning with electric data in distribution networks has been visited recently.
%For these reasons, Many works exist on topology identification.
Initial researches are based on strict assumptions. For example, \cite{ref:Cavraro2015d,ref:Sharon2012t} need the information of all switch locations and find right combinations; \cite{ref:Korres2012a,ref:Baran2009t} require the admittance matrix for a state-estimation-based learning method. These assumptions are unrealistic, since operators may not own complete information of circuit breakers and admittance matrix. Some more recent works overcome these assumptions but require data from all buses. For example, \cite{ref:Cavraro2019V} gives a relaxation for the non-convex maximum likelihood estimation with grid's voltage co-variances; \cite{ref:Deka2016p} uses conditional-independence tests of voltage among all nodes to select candidate lines;  \cite{ref:Bolognani2013i,ref:weng2017d} requires voltage magnitude of each node for calculating edge weights (e.g., mutual information) as the metric to structure learning; \cite{ref:Deka2015s,ref:Deka2015s2} analyze relationships of second moments with each nodal voltage magnitude and power injections, still requiring data from all the buses.
% * <liaoyizheng@gmail.com> 2019-01-02T06:13:54.652Z:
%
% > Therefore, some recent works abandon them. However, these works require measurements of all nodes
% some recent works overcome these assumptions but require data from all buses
%
% ^.

Nonetheless, in systems like secondary distribution grids, sensors are often limited to end-user nodes (observed nodes). One reason is that they are mainly installed for services like price controllable demand and load monitoring\cite{ref:Deka2016L}. There are also studies in recovering the network with hidden nodes. \cite{ref:Deka2016l1,ref:Deka2016L} introduce the second order statistics of observed nodes as a metric to learn the topology, but they need complete information of line impedance. \cite{ref:Park2017t} estimates line impedance and employ a graphical-learning algorithm \cite{ref:Choi2011l} for tree recovery. However, their assumption of uncorrelated-power injections is  disobeyed in the real world due to common customer behavior. This leads to a weak performance of their algorithms with realistic data. In addition, their requirement of voltage angles may not be satisfied in many grids due to limited deployment of micro PMUs.
% * <liaoyizheng@gmail.com> 2019-01-02T06:17:05.698Z:
%
% > they need the angle information that may be unavailable in many grids.
% the requirement of voltage angles may not be satisfied in many grids due to the limited deployment of micro PMU.
%
% ^.
% * <liaoyizheng@gmail.com> 2019-01-02T06:16:44.012Z:
%
% > realistic
% real-world
%
% ^.
% * <liaoyizheng@gmail.com> 2019-01-02T06:15:33.183Z:
%
% > secondary
% try to be consistent using either LV & MV or primary & seconedary
%
% ^.

In this paper, we aim at learning the structure of radial distribution grids only with end-user voltage and current data. For such purpose, we first convert the grid to a graphical model. Each leaf node has a unique path to the root so we can trace back from leaves to the root. In this process, edge weight is used as a metric to discover hidden nodes along the path. Then, recursive grouping (RG) algorithm \cite{ref:Choi2011l} is used to iteratively detect hidden nodes with edge weights of current-node subset and calculate edge weights related to the hidden nodes. %RG method can exactly recover the whole tree given exact edge weight of observed nodes.
% * <liaoyizheng@gmail.com> 2019-01-02T06:18:01.055Z:
%
% only with end-user voltage and current data
%
% ^.

A key step of RG algorithm is estimating line impedance (i.e., edge weight) of the end-node subset, which requires the nodal current deviation from the statistical mean to be pairwise uncorrelated. However, realistic data still presents low correlations, thus leading to accumulative errors. To eliminate correlations, a whitening technique is employed to make the whitened current deviations uncorrelated. With observed measurements, we prove Cholesky whitening preserves the values of whitened data due to its upper-diagonal structure. Namely, Cholesky whitening implemented on partial data gives exactly the corresponding-partial block of the whitened data from both the observed and the hidden. Therefore, an accurate estimation of line impedances from partial measurements is guaranteed.
% * <liaoyizheng@gmail.com> 2019-01-02T06:19:32.737Z:
%
% > Before RG
% A key step of RG algorithm is estimating line impedance ..., which requires the current injection at each bus to be independent
%
% ^.
% * <liaoyizheng@gmail.com> 2019-01-02T06:19:17.256Z:
%
% > mean
% statistical mean
%
% ^.

Finally, this paper increases the robustness of the learning algorithm by handling the scenarios where only measurement magnitudes are available or the grid is unbalanced. For the first scenario, we employ clustering techniques to select measurements with similar angles that can be therefore discarded. Subsequently, we utilize induction method to prove modulus of impedance with path information is still a feasible metric for RG algorithm to obtain the whole tree. For the second scenario, we propose an approximation for the Ohm's law so that our structure learning methods can work.
% * <liaoyizheng@gmail.com> 2019-01-02T06:21:20.406Z:
%
% > three-phase unbalanced
% unbalanced three-phase
%
% ^.
The performance of the method is verified on multiple IEEE distribution systems. Real-world load data and the $115$-bus grid from Pacific Gas and Electric Company (PG\&E) are also included. The result of simulated data and real-time measurements shows high accuracy of our algorithms.

The rest of the paper is organized as follows: Section \ref{section2} illustrates the impedance estimation and RG algorithm. Section \ref{section3} introduces Cholesky whitening to get rid of real-world correlations. Section \ref{section4} considers the magnitude-available case or unbalanced networks. Section \ref{section5} exhibits experiments. Section \ref{section5} makes the conclusions.
\vspace{-4mm}
\section{A graphical structure learning}
\label{section2}
% For a period of measurements, the underlying topology keeps unchanged. A graphical model is a good approach to conduct the topology recovery since it organizes measurements in the form of joint probability distribution without any preference. This section converts the radial grid to the graphical model. Edge weight (line impedance) between every two end nodes is estimated with terminal measurements. Then, we propose a graphical-learning algorithm to recursively update a node subset and identify the radial structure.

\subsection{Graphical Model of Distribution Grids}
\label{Part 2A}
A graphical model organizes measurements in the form of joint probability distribution without any preference. It is suitable for structure learning since the same topology constrains a period of data without bias. We model the radial network as $\mathcal{G}=(\mathcal{V},\mathcal{E})$. $\mathcal{V}$ denotes the node set that represents grid buses and $\mathcal{E}$ denotes the edge set for distribution lines. Among all the buses, we denote observed node set $\mathcal{O}$ to represent nodes with meters. The hidden-node set $\mathcal{H}$ includes nodes without power consumptions and lacks measurements. Such a representation is termed as a latent tree \cite{ref:Choi2011l}.
%According to \cite{ref:Choi2011l}, a hidden node with degree-$2$ lacks information diversity for any identifiable. Therefore, we focus on the following scenario:
In a latent tree, the intermediate nodes can't exhibit any identifiable information if they lie on an edge without extra branches. Thus, we focus on identifying hidden nodes with at least $3$ branches:
\begin{assumption} \label{assum1}
In our study scope, the hidden nodes have a degree larger than $2$.
\end{assumption}

\begin{figure}
\centering
\includegraphics[width=3.5in]{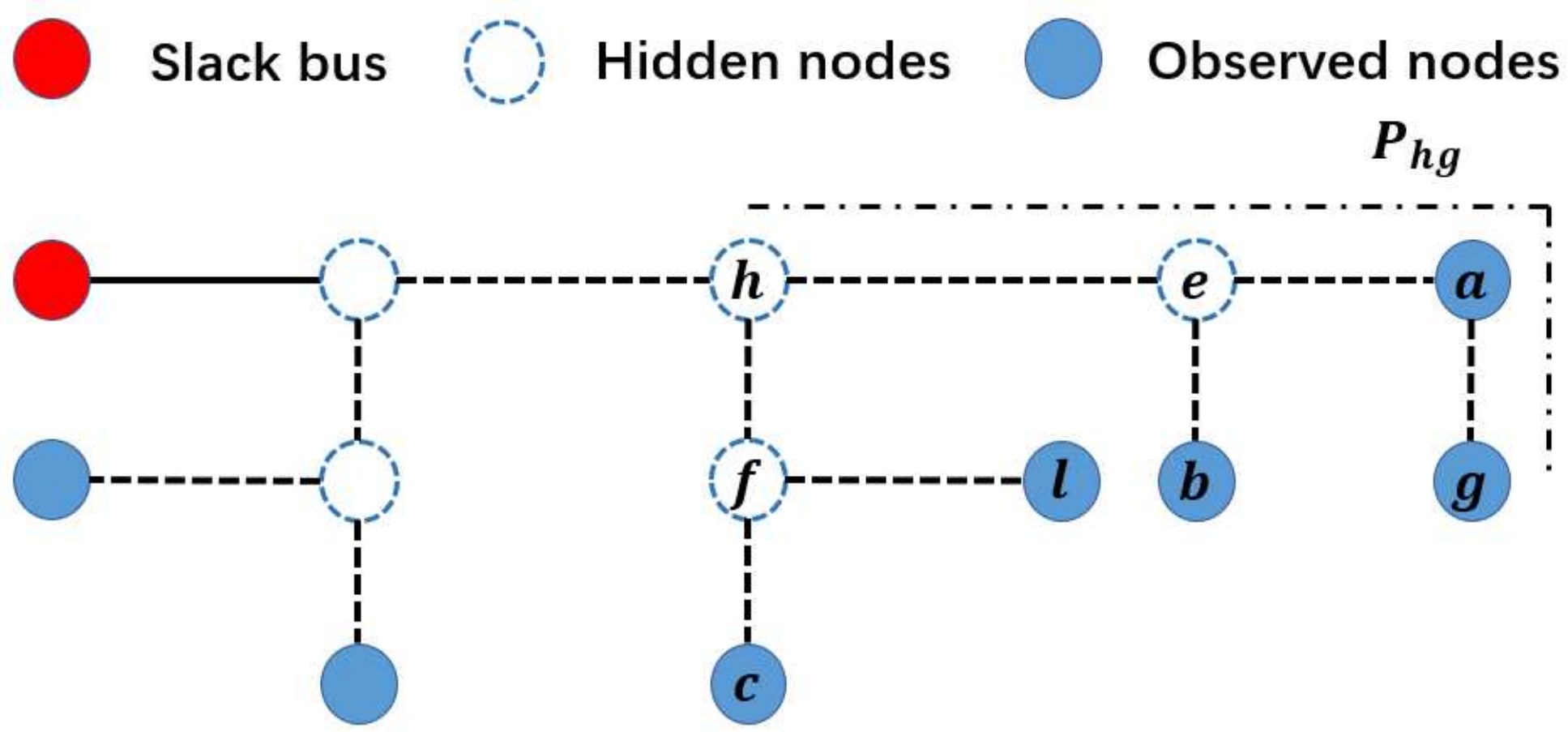}
\centering
\caption{An example of a latent tree. The root node is the slack bus and all the observed nodes consist of $\mathcal{O}$. Due to Assumption \ref{assum1}, intermediate node $a$ has $2$ degrees and only can be treated as an observed node. We introduce the following concepts for the convenience of later derivations: path $\mathcal{P}_{hg}$ is defined as the set of lines that connect node $h$ and $g$; if node $a$ lies on the path from node $g$ to the slack bus, we call node $a$ the parent of node $g$ and node $g$ the child of node $a$; we call the set of child nodes a sibling group; the sibling group of node $f$ is designated as $\mathcal{C}(f)$. Here, $\mathcal{C}(f)=\{c,l\}$.}
\label{figure_graphical_model}
\vspace{-6mm}
\end{figure}

The latent tree model with Assumption \ref{assum1} can be easily understood by Fig. \ref{figure_graphical_model}, where the root node is the slack bus and all the observed nodes consist of $\mathcal{O}$. Due to Assumption \ref{assum1}, intermediate node $a$ has $2$ degrees and only can be treated as an observed node. We introduce the following concepts for the convenience of later derivations: path $\mathcal{P}_{hg}$ is defined as the set of lines that connect node $h$ and $g$; if node $a$ lies on the path from node $g$ to the slack bus, we call node $a$ the parent of node $g$ and node $g$ the child of node $a$; we call the set of child nodes a sibling group; the sibling group of node $f$ is designated as $\mathcal{C}(f)$. Here, $\mathcal{C}(f)=\{c,l\}$.

In the graphical model for distribution grids, we treat the voltage and current injections of each bus as complex random variables represented via a collection of data. They are stored in $N\times L$ matrices $\mathbf{V}=(\mathbf{v}_1,\mathbf{v}_2,\cdots,\mathbf{v}_L)$ and $\mathbf{I}=(\mathbf{i}_1,\mathbf{i}_2,\cdots,\mathbf{i}_L)$, where $N$ is the total number of time slots, $L$ is the number of buses and $\mathbf{v}_l$ and $\mathbf{i}_l$	($1\leq l\leq L$) represent collections of voltage and current from time $1$ to $N$ for bus $l$. These measurements are constrained via Ohm's law whose parameters, i.e., line impedances, are assumed to be edge weights in the graphical model.
\vspace{-4mm}
\subsection{Edge Weight Estimation for Structure Learning}
\label{Part 2B}
%This subsection estimates the sum of line impedances in the path $\mathcal{P}_{ab}$ ($\forall a,b\in\mathcal{O}$). For a tree structure, the path between observed nodes reveals the underlying connectivity. We show in the Appendix \ref{appenA} that paths among all observed nodes are enough to recover the whole tree.
For the latent tree, additive edge weights for the path among observed nodes are enough to recover the whole tree via mathematical manipulation like summation and subtraction, which is illustrated in the recursive grouping (RG) algorithm \cite{ref:Choi2011l} in Appendix \ref{appenA}. Consequently, in this part, we illustrate how to estimate the required edge weights with voltage and current phasors from observed nodes.
% In a latent tree, we treat the voltage and current injections of each bus as complex random variables. They can be represented as $\mathbf{V}=(V_1,V_2,\cdot \cdot \cdot,V_L)$ and $\mathbf{I}=(I_1,I_2,\cdot \cdot \cdot,I_L)$, where $L$ is the number of buses. At discrete time $n$, the voltage and current measurements at bus $a$ are $v_a[n]$ and $i_a[n]$, respectively. A collection of voltage and current measurements at time $n$ are denoted as $\mathbf{v}[n]=(v_1[n],v_2[n],\cdot \cdot \cdot,v_L[n])$ and $\mathbf{i}[n]=(i_1[n],i_2[n],\cdot \cdot \cdot,i_L[n])$. Then, we use $\mathbf{v}^{1:N}=(\mathbf{v}[1],\mathbf{v}[2],\cdot \cdot \cdot,\mathbf{v}[N])^T$ and $\mathbf{i}^{1:N}=(\mathbf{i}[1],\mathbf{i}[2],\cdot \cdot \cdot,\mathbf{i}[N])^T$ to denote all voltage and current measurements from time $1$ to $N$, where $T$ is the transpose operator. For bus $l$ ($1\leq l\leq L$), its collections of voltage and current from time $1$ to $N$ are denoted as $\mathbf{v}_l=(v_l[1],v_l[2],\cdots,v_l[N])^T$ and $\mathbf{i}_l=(i_l[1],i_l[2],\cdots,i_l[N])^T$. At time $n$ and bus $l$, deviations of voltage and current from the mean are defined as $\Delta v_l[n]= v_l[n]-E[\mathbf{v}_l]$ and $\Delta i_l[n]=i_l[n]-E[\mathbf{i}_l]$, where $E[\cdot]$ represents the mean operation. Similarly, matrices $\Delta\mathbf{v}^{1:N}$ and $\Delta\mathbf{i}^{1:N}$ contain row vectors $\Delta \mathbf{v}[n]$ and $\Delta \mathbf{i}[n]$ for all buses' data at time $n$ and column vectors $\Delta \mathbf{v}_l$ and $\Delta \mathbf{i}_l$ for bus $l$ from time $1$ to $N$.

To estimate edge weights (i.e., line impedances) from data, we start from Ohm's law since it integrates impedances with voltage and current measurements. Let $\mathbf{Y}$ be the admittance matrix of the grid and the nodal network equation is $(\mathbf{I})^T=\mathbf{Y}(\mathbf{V})^T$ from time $1$ to $N$, where $T$ is the transpose operator.

To transform $\mathbf{Y}$ into the impedance matrix, we need to eliminate the column and row in $\mathbf{Y}$ with respect to the slack bus so that $\mathbf{Y}$ is invertible \cite{ref:bapat2010g}. If we consider the deviation of voltage and current from the statistical mean, Ohm's equation is still valid due to its linearity. In the following derivation, we only consider deviation data and keep the notation unchanged. Let node $1$ be the slack bus and we know $\mathbf{v}_1=\mathbf{0}_{N\times 1}$, which brings no statistical information. Thus, we can eliminate the first column of $\mathbf{V}$ and $\mathbf{I}$ and the first row and column of $\mathbf{Y}$ matrix, but the equation still holds. Without loss of generality, we use the same notations for the eliminated matrices in the following derivations. After the elimination, we denote $\mathbf{Z}=\mathbf{Y}^{-1}$. Consequently, we have $(\mathbf{V})^T=\mathbf{Z}(\mathbf{I})^T$.
% * <liaoyizheng@gmail.com> 2019-01-02T06:29:27.391Z:
%
% > $(\mathbf{i}^{1:N})^T=Y(\mathbf{v}^{1:N})^T$
% this matrix multiplication seems not right.
%
% ^.

While the target is to estimate the path impedance, Lemma \ref{lemma1}\cite{ref:Deka2016l1,ref:bapat2010g} links it to the entry of $\mathbf{Z}$.
\vspace{-1.5mm}
\begin{lemma}\label{lemma1}
In a radial distribution grid, $\mathbf{Z}(a,b)$ is the sum for line impedances of the common path between nodes $a$, $b$ to the slack bus, $a,b\in\{2,3,\cdot \cdot \cdot,L\}$ (node $1$ is the slack bus).
\end{lemma}
\vspace{-1mm}
Mathematically, we have: $\mathbf{Z}(a,b)=\sum\limits_{(ik)\in \mathcal{P}_{a1}\bigcap \mathcal{P}_{b1}}z_{ik}$, where $\mathcal{P}_{a1}$ and $\mathcal{P}_{b1}$ are paths from nodes $a$ and $b$ to the slack bus. $z_{ik}$ is the impedance of line $(ik)$. To specify each entry of $\mathbf{Z}$, we zoom in every column of $\mathbf{V}$: $\mathbf{v}_a=\sum\limits_{l=2}^{L}\mathbf{Z}(a,l)\mathbf{i}_l$ ($\forall a\in \{2,3,\cdot \cdot \cdot,L\}$).
% \begin{equation} \label{eqn:rowohm}
% \Delta\mathbf{v}_a=\sum\limits_{l=2}^{L}Z(a,l)\Delta\mathbf{i}_l,\ \ \forall a\in \{2,3,\cdot \cdot \cdot,L\}.
% \end{equation}
This equation can be left-multiplied by another current-deviation vector ${\mathbf{i}_b}^{H}$ and if current deviations are uncorrelated, we acquire the entry $\mathbf{Z}(a,b)$. Therefore, the following assumption is made to give an unrealistic impedance estimation. We show how to relax this assumption and obtain an accurate result in Section \ref{section3}.
% Fig. \ref{figure_coefficient_correlation} illustrates the cumulative distribution function (CDF) for the modulus of correlation coefficients ($|\rho|$) for the current (Fig. \ref{figure_coefficient_correlation} $(a)$)
% % ($|\rho_{I}|$)
% and current deviation
% % ($|\rho_{\Delta I}|$).
% (Fig. \ref{figure_coefficient_correlation} $(b)$). They are acquired with real-world data from PG\&E. We find that current deviations have less correlations and propose Assumption \ref{assum2}.
% * <liaoyizheng@gmail.com> 2019-01-02T06:28:09.336Z:
%
% > illustrate
% for fig 2, try to set the x-axis all from 0 to 1
%
% ^.

% \begin{figure}
% \centering
% \subfigure[Cumulative distribution function (CDF) for modulus of coefficient correlations for current. Above $80\%$ of values are distributed from $0.1$ to $0.7$.]{
% \includegraphics[width=3.4in]{Fig_correlation1}}
% \hspace{0.5in}
% \subfigure[Cumulative distribution function (CDF) for modulus of coefficient correlations for current deviations. Near 95\% of values are less than $0.15$.]{
% \includegraphics[width=3.4in]{Fig_correlation2}}
% \centering
% \caption{CDF for modulus of correlation coefficients ($|\rho|$).}
% \label{figure_coefficient_correlation}
% \vspace{-6mm}
% \end{figure}

% \begin{figure}
% \centering
% \includegraphics[width=3.5in]{Fig_correlation}
% \centering
% \caption{CDF for modulus of correlation coefficients ($|\rho|$) for the current deviation $|\rho_{\Delta I}|$. Near 95\% of $|\rho_{\Delta I}|$ is less than $0.15$.}
% \vspace{-6mm}
% \label{figure_coefficient_correlation}
% \end{figure}

\begin{assumption} \label{assum2}
\vspace{-1mm}
Current deviations of different buses are pair-wise uncorrelated in distribution grids.
\end{assumption}

Based on Assumption \ref{assum2}, we apply a inner product procedure to the above equation:
\begin{equation}
\begin{aligned} \label{eqn:innerpro1}
{\mathbf{i}_b}^{H}\mathbf{v}_a&=\sum\limits_{l=2}^{L}\mathbf{Z}(a,l){\mathbf{i}_b}^{H}\mathbf{i}_l=\mathbf{Z}(a,b){\mathbf{i}_b}^{H}\mathbf{i}_b,
\end{aligned}
\end{equation}
where $H$ represents the conjugate transpose. We claim ${\mathbf{i}_b}^{H}\mathbf{i}_l=0$ ($\forall l\neq b$) because: $1$) $\Delta\mathbf{i}_b$ is uncorrelated of $\mathbf{i}_l$ due to Assumption \ref{assum2}, $2$) the mean $E[\mathbf{i}_l]=0,\ \forall 1\leq l\leq L$. Then, ${\mathbf{i}_b}^{H}\mathbf{i}_l=N\cdot E[{\mathbf{i}_b}^H\mathbf{i}_l]=N\cdot E[\mathbf{i}_b]\cdot E[\mathbf{i}_l]=0$. With voltage and current data, \eqref{eqn:innerpro1} can find $\mathbf{Z}(a,b)$ ($a,b\in \mathcal{O}$).

Considering the property of $\mathbf{Z}(a,b)$ in Lemma \ref{lemma1}, we capture the distance $d_{ab}$, representing the total impedance of path $\mathcal{P}_{ab}$.
\begin{equation} \label{eqn:distesti1}
% * <liaoyizheng@gmail.com> 2019-01-02T06:31:26.762Z:
%
% > \begin{equation} \label{eqn:distesti1}
% > d_{ab}=Z(a,a)+Z(b,b)-2Z(a,b), \ \ \forall a,b\in \mathcal{O}.
% > \end{equation}
%
% this equation is very important, try to put it into a collary or property seciton so we can highlight it
%
% ^.
d_{ab}=\mathbf{Z}(a,a)+\mathbf{Z}(b,b)-2\mathbf{Z}(a,b), \ \ \forall a,b\in \mathcal{O}.
\end{equation}
\vspace{-8mm}
\subsection{Edge Weight-based Structure Learning}
The obtained distance in \eqref{eqn:distesti1} consists of line impedance, a physically additive weight for the edge in the latent tree. For example, in Fig. \ref{figure_graphical_model}, we have $d_{eg}=d_{ea}+d_{ag}$. With additive distances among observed nodes, many methods can conduct structure learning, e.g., Recursive Grouping (RG), Neighbor Joining (NJ) and Quartet-based Methods\cite{ref:Choi2011l}. We introduce RG algorithm in Appendix \ref{appenA}.

The whole algorithm is defined as $\mathcal{G}=\mathcal{RG}(\mathbf{D},\mathcal{O})$, where $\mathbf{D}$ is the distance matrix in Appendix \ref{appenA} and Algorithm \ref{alg2} describes the structure learning process.

\begin{algorithm}
\caption{Line Impedance-based Structure Learning}
\label{alg2}
\begin{algorithmic}[1]
\REQUIRE The current and voltage measurements of observed nodes $\mathcal{O}$.
\ENSURE The structure of the latent tree.\\
\STATE Calculate ${\mathbf{i}_b}^{H}\mathbf{v}_a$ and ${\mathbf{i}_b}^{H}\mathbf{i}_b$, $\forall a,b\in \mathcal{O}$.
\STATE \textbf{for} $\forall a,b\in \mathcal{O}$, \textbf{do}
% * <liaoyizheng@gmail.com> 2019-01-02T06:33:07.241Z:
%
% > \STATE \textbf{for} $a,b\in \mathcal{O}$, \textbf{do}
% use forall command for for loop in alg env
%
% ^.
\STATE $\mathbf{Z}(a,b)={\mathbf{i}_b}^{H}\mathbf{v}_a/{\mathbf{i}_b}^{H}\mathbf{i}_b$.
\STATE $d_{ab}=\mathbf{Z}(a,a)+\mathbf{Z}(b,b)-2\mathbf{Z}(a,b)$.
\STATE \textbf{end for}
\STATE Form the estimated distance matrix $\widehat{\mathbf{D}}$ of observed nodes.
\STATE $\mathcal{G} \leftarrow \mathcal{RG}(\widehat{\mathbf{D}},\mathcal{O})$.
\end{algorithmic}
\end{algorithm}

Since the estimated distance $\widehat{\mathbf{D}}$ may have errors, we employ the following thresholds \cite{ref:Choi2011l} to relax the equality in ``Node Grouping" in Appendix \ref{appenA}. For every $k\in \mathcal{V}\backslash\{i,j\}$, if $\widehat{\Phi}_{ijk}-\hat{d}_{ij}<\epsilon_1$, $j$ is the parent of $i$. If $\Psi_{ij}=\max\limits_{k\in \mathcal{V}\backslash\{i,j\}}\widehat{\Phi}_{ijk}-\min\limits_{k\in \mathcal{V}\backslash\{i,j\}}\widehat{\Phi}_{ijk}<\epsilon_2$, $i$ and $j$ are siblings. The thresholds $\epsilon_1$ and $\epsilon_2$ are pre-defined constants.
% * <liaoyizheng@gmail.com> 2019-01-02T06:34:15.791Z:
%
% > For every $k\in \mathcal{V}\backslash\{i,j\}$, if $\widehat{\Phi}_{ijk}-\hat{d}_{ij}<\epsilon_1$, $j$ is the parent of $i$. If $\Psi_{ij}=\max\limits_{k\in \mathcal{V}\backslash\{i,j\}}\widehat{\Phi}_{ijk}-\min\limits_{k\in \mathcal{V}\backslash\{i,j\}}\widehat{\Phi}_{ijk}<\epsilon_2$, $i$ and $j$ are siblings. The thresholds $\epsilon_1$ and $\epsilon_2$ are pre-defined constants.
% use new line of equations to describe it
%
% ^.

\vspace{-3mm}
\section{Enhanced structure learning with measurement decorrelation}
\label{section3}

\begin{figure}
\centering
\includegraphics[width=3.5in]{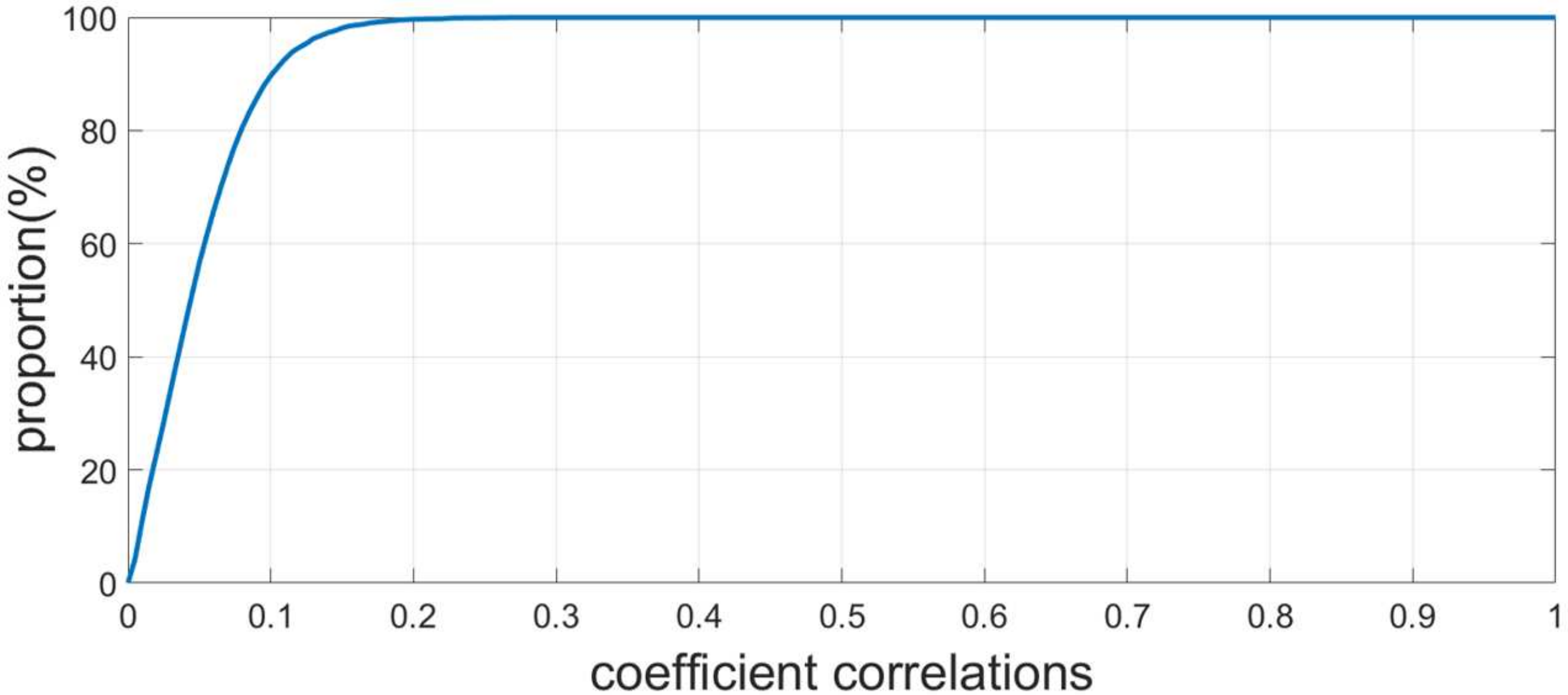}
\centering
\caption{CDF for modulus of correlation coefficients ($|\rho|$) for the current deviation. There are correlations among different nodes.}
\vspace{-6mm}
\label{figure_coefficient_correlation}
\end{figure}

While the above learning process employs the uncorrelation requirement for current deviations in Assumption \ref{assum2}, the load-current deviations in real world usually present correlations due to similar power consumption patterns. For instance, Fig. \ref{figure_coefficient_correlation} illustrates the cumulative distribution function (CDF), obtained from real-world data in PG\&E, for modulus of correlation coefficients ($|\rho|$) for the load-current deviation. Therefore, the error term $\textbf{i}_b^H \textbf{i}_l$ ($\forall l\neq b, \ l,b\in \mathcal{O}$) in \eqref{eqn:innerpro1} are nonzeros. Meanwhile, the coefficients $\mathbf{Z}(a,l)$ ($\forall a\in \mathcal{O}$) in \eqref{eqn:innerpro1} are mostly nonzeros. The accumulative non-zero terms may lead to large estimation errors and non-robust learning process, which is shown numerically in Section \ref{Section4B}.

On the other hand, the hidden nodes usually function as power-separation nodes. They have near-zero power consumptions and don't need meters. Therefore, we can assume their current deviations are uncorrelated with those of observed nodes. Hence we consider the following scenario:
\vspace{-1.5mm}
\begin{assumption} \label{assum3}
Only current deviations of observed nodes have statistical correlations.
\end{assumption}
\vspace{-1.5mm}

% We employ the whitening matrix to map these current-change vectors to the linearly independent space. The inversely mapping works on the $Z$ matrix to satisfy Ohm's law. With independences of mapped-current changes, the entry of the inversely mapped $Z$ matrix is obtained via inner product procedure. To estimate the original $Z$ matrix, we only consider diagonal elements of the inverse whitening matrix. We prove this estimation process performs well with Cholesky whitening matrix because of its special structure.

% The off-diagonal elements of the inverse whitening matrix bring about errors. Part of them are observable from the observed measurements while others are unobservable. In Section \ref{Section3C}, the information of observable parts is exploited in the distance estimation to lower the errors. Combing this estimation process with the RG algorithm, we propose a enhanced structure learning method. We prove this method works well in small grids with high share of observed nodes.

% In Section \ref{section4}, this learning method performs relatively bad in large-size distribution grid with small percentage of observed nodes. To increase the account for observed nodes, this paper provides a network reduction (NR) technique. The key idea is to create an artificial node in place of many unobserved nodes through a wise placement of smart sensors. Finally, we put forward a NR-based structure learning flowchart to deal with large systems. This learning methodology is validated in the later numerical test.
\vspace{-4mm}
\subsection{Whitening in Impedance Estimation}
\label{Section3A}

Under Assumption \ref{assum3}, this subsection adopts whitening transformation to eliminate current-deviation correlations for observed nodes. Normally, whitening is conducted via a whitening matrix $\mathbf{W}$, which can be directly calculated from current-deviation data. The whitening process is as follows: $(\tilde{\mathbf{I}})^T=\mathbf{W}(\mathbf{I})^T=\mathbf{W}\mathbf{Y}(\mathbf{V})^T$, where $\tilde{\mathbf{I}}$ is the whitened measurements with orthogonal column vectors.
% \begin{equation}  \label{eqn9}
% \Delta\tilde{\mathbf{i}}^{1:N}=W\Delta\mathbf{i}^{1:N}=WY\Delta\mathbf{v}^{1:N}.
% \end{equation}

% The original $(L-1)\times N$ ($L$ is the number of nodes and $N$ is the time period) matrix $\Delta\mathbf{i}^{1:N}$ is row full rank as long as we sample enough data to make sure $N\gg L-1$. Therefore, the $(L-1)\times (L-1)$ whitening matrix $W$ exists in order that $\Delta\tilde{\mathbf{i}}^{1:N}$ has orthogonal row vectors.

The eliminated admittance matrix $\mathbf{Y}$ is invertible and whitening transformations like Cholesky whitening and zero-phase component (ZCA) whitening are invertible \cite{ref:kessy2018optimal}. To acquire $\mathbf{Z}$ matrix, we reach to the whitened Ohm's equation: $(\mathbf{V})^T=\mathbf{Z}\mathbf{M}(\tilde{\mathbf{I}})^T=\mathbf{K}(\tilde{\mathbf{I}})^T$,
% \begin{equation} \label{eqn10}
% \Delta\mathbf{v}^{1:N}=ZM\Delta\tilde{\mathbf{i}}^{1:N}=K\Delta\tilde{\mathbf{i}}^{1:N},
% \end{equation}
where we denote $\mathbf{K}=\mathbf{Z}\mathbf{M}$ and $\mathbf{M}$ is the inverse of $\mathbf{W}$ matrix.

However, we can only obtain an observed whitening matrix $\widehat{\mathbf{W}}_O$ with observed current-deviation measurements. We utilize the notation $\widehat{\cdot}$ to represent matrix calculated from the observed measurements and the subscript $O$ to represent the sub-matrix corresponding to observed nodes in the matrix of all nodes. For example, usually $\widehat{\mathbf{W}}_O\neq \mathbf{W}_O$ since $\widehat{\mathbf{W}}_O$ is only obtained from observed measurements:
\begin{equation} \label{eqn:obserwhite}
(\widehat{\tilde{\mathbf{I}}}_O)^T=\widehat{\mathbf{W}}_O(\mathbf{I}_O)^T,
\end{equation}
where $\mathbf{I}_O$ and $\widehat{\tilde{\mathbf{I}}}_O$ denote the observed current deviations and whitened-observed current deviations. In Section \ref{Section3B}, Theorem \ref{theo2}, we will prove that under Cholesky whitening, $\widehat{\tilde{\mathbf{I}}}_O=\tilde{\mathbf{I}}_O$. Therefore, we apply the inner product to obtain entry of $\mathbf{K}$: $\mathbf{K}_O(a,b)={\tilde{\mathbf{i}}_b}^{H}\mathbf{v}_a/{\tilde{\mathbf{i}}_b}^{H}\tilde{\mathbf{i}}_b$, ($\forall a,b\in \mathcal{O}$),
% \begin{equation} \label{eqn12}
% K_O(a,b)=\frac{\Delta\mathbf{v}_a{\Delta\tilde{\mathbf{i}}_b}^{H}}{\Delta\tilde{\mathbf{i}}_b{\Delta\tilde{\mathbf{i}}_b}^{H}},\ \forall a,b\in \mathcal{O},
% \end{equation}
where ${\tilde{\mathbf{i}}_b}$ is the $b^{th}$ column vector of $\tilde{\mathbf{I}}$ and $\mathbf{K}_O$ is the observed block of $\mathbf{K}$ matrix. Then, we give an estimation for the observed block of $\mathbf{Z}$ matrix, denoted as $\widehat{\mathbf{Z}}_O$:
\vspace{-1.5mm}
\begin{equation} \label{eqn:Zestimation}
\widehat{\mathbf{Z}}_O=\mathbf{K}_O\widehat{\mathbf{M}}_O^{-1}=\mathbf{K}_O\widehat{\mathbf{W}}_O.
\end{equation}In the following subsection, we will prove Cholesky whitening can offer a relatively accurate estimation with \eqref{eqn:Zestimation}.
% \begin{equation} \label{eqn13}
% \widehat{Z}(a,b)=\frac{K(a,b)}{\widehat{M}(b,b)},\ \forall a,b\in \mathcal{O}.
% \end{equation}

Unlike $\mathbf{Z}$ matrix, the estimated impedance matrix $\widehat{\mathbf{Z}}_O$ is asymmetric. The estimated distance between two observed nodes $\forall a,b\in \mathcal{O}$ can be gained as follows:
% * <liaoyizheng@gmail.com> 2019-01-02T06:36:16.421Z:
%
% >  $Z$
% all matrix notation should be bold math
%
% ^.
\vspace{-1.5mm}
\begin{equation}
\label{eqn:distesti2}
\hat{d}_{ab}=\widehat{\mathbf{Z}}_O(a,a)+\widehat{\mathbf{Z}}_O(b,b)-\widehat{\mathbf{Z}}_O(a,b)-\widehat{\mathbf{Z}}_O(b,a).
\end{equation}

% \begin{remark}
% The whitening-based distance estimation is a generalization of estimation methodology in Section \ref{section2}. When the whitening matrix $W=I$, the whitening based procedure will devolve into the method in Section \ref{section2}.
% \end{remark}
\vspace{-4mm}
\subsection{Choice of Whitening: Cholesky Whitening}
\label{Section3B}
We prove Cholesky whitening gives an accurate result via the above estimation process. Generally, a whitening matrix $\mathbf{W}$ aims at transforming the covariance matrix to be the identity matrix, then we have: $\mathbf{W}^H\mathbf{W}=\mathbf{\Sigma}^{-1}$,
% \begin{equation} \label{eqn15}
% W^HW=\Sigma^{-1},
% \end{equation}
where $\mathbf{\Sigma}$ is the covariance matrix of $\mathbf{I}$. According to Cholesky decomposition, we know: $\mathbf{U}^H\mathbf{U}=\mathbf{\Sigma}^{-1}$,
% \begin{equation} \label{eqn16}
% U^HU=\Sigma^{-1},
% \end{equation}
 where $\mathbf{U}$ is the unique upper diagonal matrix. Combining these two equations, we obtain: $\mathbf{M}=\mathbf{W}^{-1}=\mathbf{U}^{-1}$, and $\mathbf{M}$ is also an upper diagonal matrix. Then, we claim the following theorem.
\begin{theorem} \label{theo2}
If we arrange bus numbers in the observed nodes set $\mathcal{O}$ from $k$ to $L$ ($|\mathcal{O}|=L-k+1$), under Cholesky whitening,  $\widehat{\mathbf{M}}_O=\mathbf{M}_O$, $\widehat{\mathbf{W}}_O=\mathbf{W}_O$, and $\widehat{\tilde{\mathbf{I}}}_O=\tilde{\mathbf{I}}_O$.
\end{theorem}

The proof can be seen in Appendix \ref{appenB}. Based on Theorem \ref{theo2}, we know the calculated $\mathbf{K}_O$ matrix in \ref{Section3A} is the same as the corresponding block in $\mathbf{K}$. To derive the error term, we rewrite $\mathbf{K}=\mathbf{Z}\mathbf{M}$ in the following form:

\begin{equation} \label{eqn:blockk1}
\begin{aligned}
\begin{bmatrix}
\mathbf{K}_1 &\mathbf{K}_2\\ \mathbf{K}_3 &\mathbf{K}_O
\end{bmatrix}
=
\begin{bmatrix}
\mathbf{Z}_1 &\mathbf{Z}_2\\ \mathbf{Z}_3 &\mathbf{Z}_O
\end{bmatrix}
\cdot
\begin{bmatrix}
\mathbf{M}_1 &\mathbf{M}_2 \\ \ &{\mathbf{M}}_O
\end{bmatrix}_{\textstyle ,}
\end{aligned}
\end{equation}
where $\mathbf{M}_O=\widehat{\mathbf{M}}_O$ by Theorem \ref{theo2}. Thus, $\mathbf{K}_O$ can be written next:
\begin{equation} \label{eqn:blockk2}
\mathbf{K}_O=\mathbf{Z}_3\mathbf{M}_2+\mathbf{Z}_O\widehat{\mathbf{M}}_O.
\end{equation}

Consequently, the distance-estimation error comes from off-diagonal block ($\mathbf{M}_2$) of the $\mathbf{M}$ matrix. The following theorem claims that under Assumption \ref{assum3}, we have $\mathbf{M}_2= \mathbf{0}_{(k-1)\times (L-k+1)}$.
\begin{theorem} \label{theo3}
If current-deviation correlations only exist among observed nodes, the off-diagonal block in the inverse of Cholesky whitening matrix (i.e., $\mathbf{M}_2$ in $\mathbf{M}$ matrix) is a zero matrix.
\end{theorem}

The proof can be seen in Appendix \ref{appenC}. Theorem \ref{theo3} claims the feasibility of $\mathbf{Z}$ matrix estimation in \eqref{eqn:Zestimation}. Finally, we come to our structure learning Algorithm \ref{alg3}.
\begin{algorithm}
\caption{Cholesky Whitening-based Structure Learning}
\label{alg3}
\begin{algorithmic}[1]
\REQUIRE The current and voltage measurements of observed nodes $\mathcal{O}$.
\ENSURE Structure of the latent tree and line parameter estimations.\\
\STATE Whiten the current deviations of the observed nodes: $\tilde{\mathbf{I}}_{O}=\widehat{\mathbf{W}}_O\mathbf{I}_{O}$.
\STATE Calculate ${\tilde{\mathbf{i}}_b}^{H}\mathbf{v}_a$ and ${\tilde{\mathbf{i}}_b}^{H}\tilde{\mathbf{i}}_b,\ \forall a,b\in \mathcal{O}$.
\STATE \textbf{for} $\forall a,b\in \mathcal{O}$, \textbf{do}
\STATE $\mathbf{K}(a,b)={\tilde{\mathbf{i}}_b}^{H}\mathbf{v}_a/{\tilde{\mathbf{i}_b}}^{H}\tilde{\mathbf{i}}_b$.
\STATE $\widehat{\mathbf{Z}}_O=\mathbf{K}_O\widehat{\mathbf{W}}_O$.
\STATE $\hat{d}_{ab}=\widehat{\mathbf{Z}}_O(a,a)+\widehat{\mathbf{Z}}_O(b,b)-\widehat{\mathbf{Z}}_O(a,b)-\widehat{\mathbf{Z}}_O(b,a),\ \forall a,b\in \mathcal{O}$.
\STATE \textbf{end for}
\STATE Use the distance $\hat{d}_{ab},\ \forall a,b\in \mathcal{O}$ to form the estimated distance matrix $\widehat{\mathbf{D}}$ of observed nodes.
\STATE $\mathcal{G}=\{\mathcal{V},\mathcal{E}\}$ $\leftarrow$ $\mathcal{RG}(\widehat{\mathbf{D}},\mathcal{O})$.
\end{algorithmic}
\end{algorithm}

\vspace{-6mm}

\section{Structure learning without angle or in an unbalanced network}
\label{section4}
In addition to the correlation issue, there are other realistic challenges for the above learning process: $1$) only voltage and current magnitudes are available in many grids, and $2$) many distribution grids are unbalanced. In this subsection, we propose methods to address these two challenges and further increase the robustness of the learning process.
\vspace{-4mm}
\subsection{Data Selection to Eliminate Angle Information}
\label{sec:data_clustering}
For challenge $1$), the absence of angle information forces us to select proper measurements so that the impact of angle differences is reduced in the structure learning. We consider adding modulus in equation \eqref{eqn:innerpro1}: $|\mathbf{Z}(a,b)|=|{\mathbf{v}_a{\mathbf{i}_b}^{H}}/{\mathbf{i}_b{\mathbf{i}_b}^{H}}|,\ \forall a,b\in \mathcal{O}$, where $|\cdot|$ represents the modular operation to a complex number or the element-wise modular operation to a complex vector or matrix. To eliminate angle information in the right-hand side, the following assumption is introduced:
\begin{assumption}\label{assum4}
In a collection of data, the angles of voltage and current are nearly unchanged in distribution grids.
\end{assumption}

Assumption \ref{assum4} can be achieved via carefully picking up of data. The voltage is usually stable when the topology is unchanged. The abrupt change of voltage magnitudes indicates the change of the topology. In addition, we prefer data with a smooth nodal-reactive-power change that brings an ignorable change for current angles. Rather than segmentation for a continuous time period, clustering techniques are employed to select data from the historical dataset. For time slot $n$, we map the sampled data into a high dimensional space $z[n]=(|\mathbf{v}_O[n]|,\lambda \cdot\mathbf{q}_O[n])$, where $|\mathbf{v}_O[n]|$ and $\mathbf{q}_O[n]$ represent the collection of voltage magnitudes and nodal-reactive-power deviations from the mean for observed nodes in time $n$, respectively, and $\lambda$ is a weight term. The reactive power deviation is utilized for normalization, but the variance information is preserved. Consequently, we use methods like k-means or hierarchical clustering to form clusters in the mapped space with Euclidean distance as the metric. When the real-time point comes, the cluster to which the new point belongs offers an appropriate data collection for Assumption \ref{assum4}.

Based on Assumption \ref{assum4}, angles can be totally eliminated and we have $|\mathbf{Z}(a,b)|={|{\mathbf{i}_b}|^{H}}\cdot|\mathbf{v}_a|/(|\mathbf{i}_b|\cdot|{\mathbf{i}_b}|^{H})$. If we consider the whitening process, we can obtain the approximated $|\mathbf{K}_O|$ and $|\mathbf{W}_O|$ matrices from voltage and current magnitudes. Similar to \eqref{eqn:Zestimation}, we give an approximation for $|\mathbf{Z}_O|=|\mathbf{K}_O|\cdot|\widehat{\mathbf{M}}_O^{-1}|=|\mathbf{K}_O|\cdot|\widehat{\mathbf{W}}_O|$. We will show this is a good approximation in the numerical experiments Section \ref{Section4C}.
\vspace{-5mm}
\subsection{Structure Learning without Angles}
Since the modulus $|\mathbf{Z}(a,a)|$ obtained above represents the modulus of sum impedances along the path from node $a$ to the slack bus, the availability of each path from leaf nodes to the root suggests the possibility to recover the tree.

With the modulus of components in $\mathbf{Z}$ matrix, we give a distance estimation for observed nodes $a$ and $b$: $d_{ab}=|\mathbf{Z}(a,a)|+|\mathbf{Z}(b,b)|-2|\mathbf{Z}(a,b)|$. Though $d_{ab}$ can't represent the true impedance between nodes $a$ and $b$ now, we use induction method to prove that it helps the RG algorithm to recover the tree. The proof can be seen in Appendix \ref{appenD}.

\vspace{-4mm}
\subsection{Structure Learning in An Unbalanced Network}
For unbalanced networks in challenge $2$), we give a general form to approximate Ohm's law via the impedance model of four-wire (with neural conductor) in Fig. \ref{figure_4_wire_model} \cite{Ebrahimi2011E}. We use the corresponding lowercase to represent phase $A$, $B$, $C$, $N$ and the ground $G$, $G^{'}$:

\begin{figure}
\centering
\includegraphics[width=3.5in]{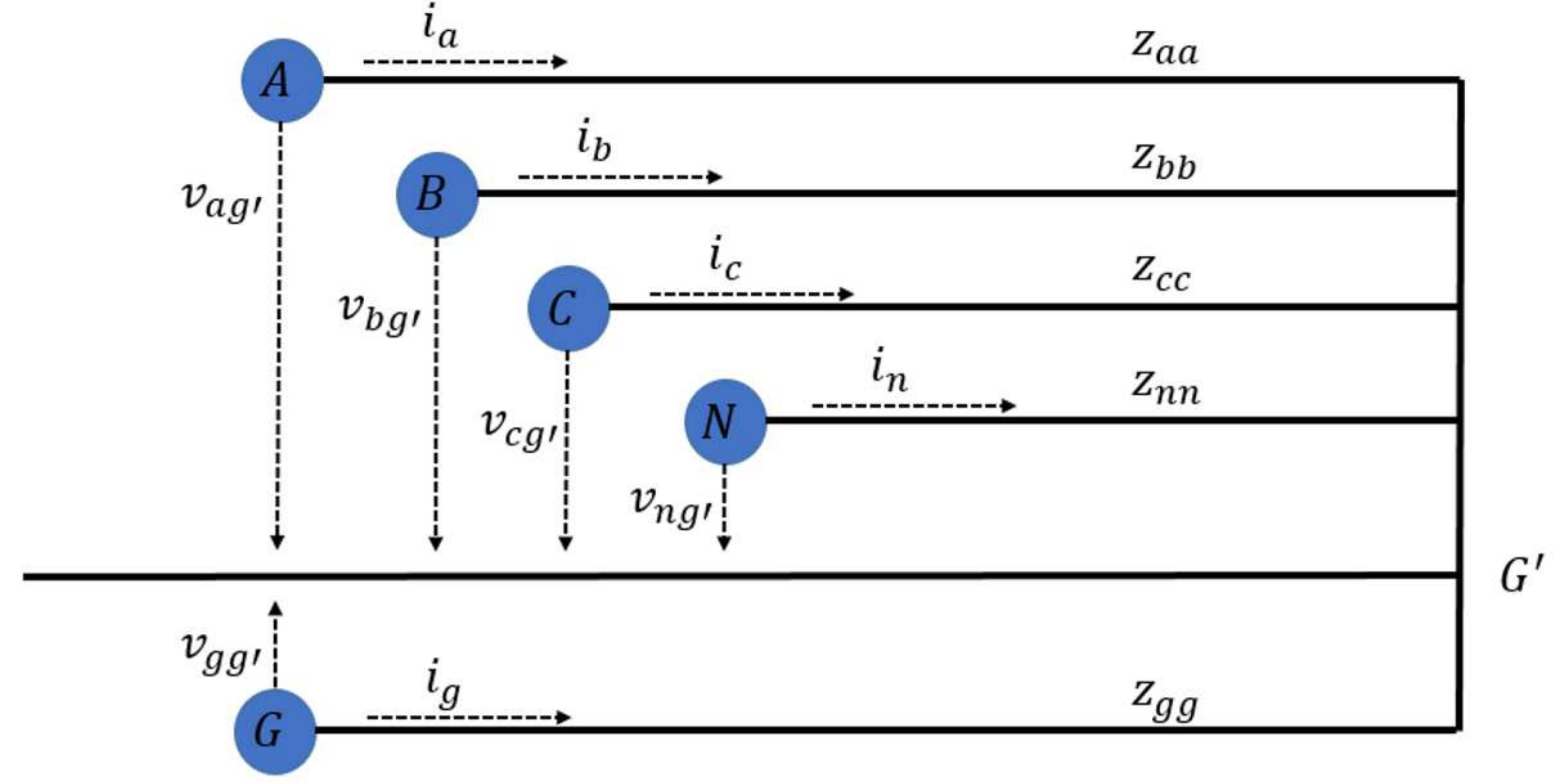}
\centering
\vspace{-4mm}
\caption{The four-wire impedance model.}
\vspace{-4mm}
\label{figure_4_wire_model}
\end{figure}
\vspace{-3mm}

\begin{equation}
\label{eqn:four_wire_voltage}
\begin{aligned}
\begin{bmatrix}
v_{ag^{'}}\\ v_{bg^{'}} \\v_{cg^{'}} \\v_{ng^{'}} \\v_{gg^{'}}
\end{bmatrix}
=
\begin{bmatrix}
z_{aa} &z_{ab} &z_{ac} &z_{an} &z_{ag}\\ z_{ba} &z_{bb} &z_{bc} &z_{bn} &z_{bg}\\ z_{ca} &z_{cb} &z_{cc} &z_{cn} &z_{cg}\\z_{na} &z_{nb} &z_{nc} &z_{nn} &z_{ng} \\z_{ga} &z_{gb} &z_{gc} &z_{gn} &z_{gg}
\end{bmatrix}
\cdot
\begin{bmatrix}
i_{a}\\ i_{b} \\i_{c} \\i_{n} \\i_{g}
\end{bmatrix}_{\textstyle ,}
\end{aligned}
\end{equation}
where $v_{kg^{'}}$, ($\forall k\in\{a,b,c,n,g\}$) represents the nodal phase voltage to the ground, $i_k$ represent the phase current to the ground, $z_{kk}$ represents the self impedance, and $z_{ik}$ ($\forall i\neq k,\ i,k\in \{a,b,c,n,g\}$) represents the mutual impedance. Carson assumes the sum of the wire current returns from the ground $i_g=\sum\limits_{k\in \{a,b,c,n,g\}}i_k$, analyzes their electromagnetic relations and give the unit self and mutual impedance for each wire and the ground \cite{Ebrahimi2011E}. We can assume we have the prior information for the impedance ratio. In phase A, we write $\lambda_{ak}=z_{ak}/z_{aa},\ \forall k\in \{a,b,c,n,g\}$. If phase $k$ is missing, $\lambda_{ak}=0$. Thus, we have
\vspace{-1.5mm}
\begin{equation}
\label{eqn:equivalent_ohm}
v_{ag^{'}}=z_{aa}\sum\limits_{k\in\{a,b,c,n,g\}}\lambda_{ak}i_k
=z_{aa}i_a^{'},
\end{equation}
where $i_a^{'}$ is the weighted current in phase A and \eqref{eqn:equivalent_ohm} is the equivalent Ohm's equation. Similar process can be conducted for other phases. Due to the linearity of \eqref{eqn:equivalent_ohm}, the weighted current can be extended to the nodal current injection and the network Ohm's equation still holds. Therefore, for each time slot and for each node, we utilize equation \eqref{eqn:equivalent_ohm} to acquire the nodal phase voltage and the weighted current, which form the phase voltage matrix $\mathbf{V}^{'}$ and equivalent current matrix $\mathbf{I}^{'}$. Finally, our structure learning method can be conducted with observed measurements $\mathbf{V}^{'}_O$ and $\mathbf{I}^{'}_O$.
\vspace{-3mm}
\section{Numerical experiments}
% * <liaoyizheng@gmail.com> 2019-01-02T06:38:11.484Z:
%
% > experiments
% simulations
%
% ^.
\label{section5}
We test simulated data and real-world power data from Pacific Gas and Electric Company (PG\&E) on IEEE distribution $8$-, $19$-, $33$-bus and PG\&E $115$-bus systems. In addition, we test the three-phase unbalanced case ($123$-bus system from GridLABD) with realistic three-phase data from PG\&E. For simulated data, we consider independent Gaussian distribution to generate current deviations: $\Delta \mathbf{I}\sim \mathcal{N}(\mu,\Sigma)$, where $\mu=[0,\cdot \cdot \cdot,0]^T$ and $\Sigma$ is a diagonal matrix with all diagonal elements to be $0.025$. In addition, the current injections for the hidden nodes are zero. We consider $N=8760$ samples to represent one-year data. PG\&E load profile contains hourly real power consumption of $123,000$ residential loads in North California, USA \cite{ref:weng2017d}. As for reactive power $q_i$ at bus $i$, we consider a random power factor at time $n$, $pf_i(n)\sim Unif(0.85,0.95),\ \forall n\in\{1,2,\cdot \cdot \cdot,8760\}$. Then, we have: $q_i(n)=p_i(n)\sqrt{1-pf_i(n)^2}/pf_i(n)$. We assume the hidden nodes don't have power consumptions and input the real-world data into the AC power flow solver in MATPOWER to obtain the voltage and current phasor.

\begin{figure}
\centering
\includegraphics[width=3.5in]{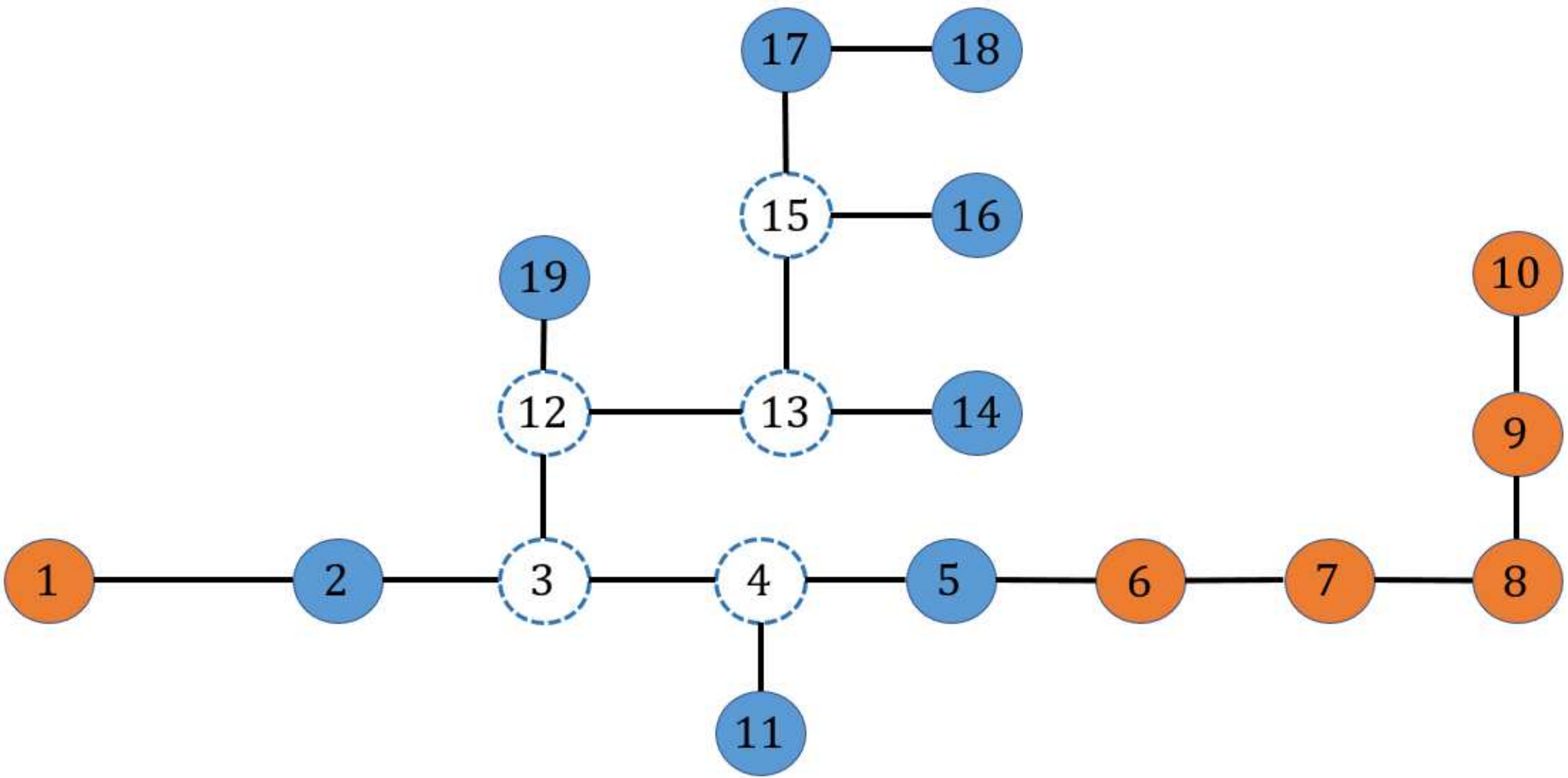}
\centering
\caption{The topology of the $19$-bus system. When $nh=5$, we assume the orange nodes are known and consider the partial topology with blue nodes and white nodes. The blue nodes are observed nodes and the white nodes are hidden nodes.}
\label{figure_19bus}
\vspace{-6mm}
\end{figure}

We define the average hidden-nodes-recovery rate ($ah(\%)$) and average correct-connection-recovery rate ($ac(\%)$) to weigh the performance of the algorithm. For a fixed number of hidden nodes ($hn$), $ah=100\times\sum\limits_{k=1}^K(N_h^{rk}/N_h^{tk})/K$, $ac=100\times\sum\limits_{k=1}^K(N_c^{rk}/N_c^{tk})/K$,
% \begin{equation*}
% \begin{aligned}
% ah&=100\times \left(\left(\frac{N_h^{r}}{N_h^{t}}\right)^{C_1}+\cdot \cdot \cdot+\left(\frac{N_h^{r}}{N_h^{t}}\right)^{C_k}\right)/K,\\
% ac&=100\times \left(\left(\frac{N_c^{r}}{N_c^{t}}\right)^{C_1}+\cdot \cdot \cdot+\left(\frac{N_c^{r}}{N_c^{t}}\right)^{C_k}\right)/K,
% \end{aligned}
% \end{equation*}
where $N_h^{rk}$, $N_h^{tk}$, $N_c^{rk}$ and $N_c^{tk}$ are the number of recovered hidden nodes, true hidden nodes, recovered connections and true connections, respectively. $K$ is the total combinations of $hn$ hidden nodes. For example, Fig. \ref{figure_19bus} is the $19$-bus system modified from the IEEE $18$-bus distribution. If we want the calculate $ah$ and $ac$ versus $hn=3$ in Fig. \ref{figure_19bus}, we should consider different combinations of $3$ hidden nodes: $(3,4,12)$, $(3, 12, 13)$ and $(12, 13, 15)$. Accordingly, we input measurements of neighboring nodes (e.g., $2$, $5$, $11$, $13$ and $19$ for the combination $(3,4,12)$) as the observed nodes.

\vspace{-4mm}
\subsection{General Performance for Balanced and Unbalanced Grids}
\label{Section4A}
This subsection presents the general performance of Cholesky whitening-based structure learning with real-world data. The input data can be phasor data or magnitude data. For the magnitude data, we show how to select the right input collection in Part \ref{Section4C}.
\begin{table}[h!]
\centering
\begin{tabular}{lcccccc}
\hline\hline
$nh$ &$\ $ &$1$ &$2$ &$3$ &$4$ &$5$\\ \hline\hline
Input phasor&$ah(\%)$  &$100$ &$100$ &$100$ &$100$ &$100$\\
$\ $ &$ah(\%)$ &$100$ &$100$ &$100$ &$100$ &$100$\\
\hline
% runtime(ms) &$0.44$ &$0.80$ &$1.2$ &$2.1$ &$3.8$\\
Input magnitude&$ah(\%)$  &$100$ &$100$ &$80.56$ &$75$ &$80$\\
$\ $ &$ac(\%)$  &$100$ &$100$ &$80.38$ &$77.78$ &$72.72$\\
% runtime(ms) &$0.42$ &$0.75$ &$1.2$ &$2.1$ &$3.8$\\
\hline\hline
\end{tabular}
\caption{\label{tab_19bus}Average recovery rate and runtime in $19$-bus system}
\vspace{-5mm}
\end{table}
For $19$-bus system in Fig. \ref{figure_19bus}, we consider different hidden node combinations and obtain Table \ref{tab_19bus}. Further, we test different systems: $8$-, $19$-, $33$-, $115$- and $123$-bus networks with $nh=1,2,3$. The $123$-bus network is unbalanced. The result is shown in Table \ref{tab_different_system}. The blank data means there is no complete combination for that case. The structure learning algorithm finds the correct connectivities with angle information in balanced networks, but obtains relatively lower accuracy with only magnitude input or in the unbalanced network. This is reasonable due to impedance-approximation error.
\begin{table}[h!]
\centering
\begin{tabular}{lccccccc}
\hline\hline
Network size &$\ $  &$8$ &$19$ &$33$ &$115$ &$\mathbf{123}$ \\ \hline\hline
$nh=1$ &$ah(\%)$  &$100$ &$100$ &$100$ &$100$ &$96.77$\\
Input phasor &$ac(\%)$  &$100$ &$100$ &$100$ &$100$ &$96.77$\\
% $\ $ &runtime(ms) &$0.42$ &$0.39$ &$0.58$ &$0.73$ &$0.82$\\
\hline
$nh=2$ &$ah(\%)$  &$\ $ &$100$ &$\ $ &$100$ &$97.81$ \\
Input phasor&$ac(\%)$  &$\ $ &$100$ &$\ $ &$100$ &$95.35$ \\
% $\ $ &runtime(ms) &$\ $ &$0.83$ &$\ $ &$2.1$ &$1.9$ \\
\hline
$nh=3$ &$ah(\%)$  &$\ $ &$100$ &$\ $ &$\ $ &$96.49$ \\
Input phasor &$ac(\%)$  &$\ $ &$100$ &$\ $ &$\ $ &$98.73$ \\
% $\ $ &runtime(ms) &$\ $ &$1.5$ &$\ $ &$\ $ &$1.7$ \\
\hline
$nh=1$ &$ah(\%)$  &$100$ &$100$ &$100$ &$100$ &$97.32$\\
Input magnitude &$ac(\%)$  &$100$ &$100$ &$100$ &$100$ &$97.25$\\
\hline
$nh=2$ &$ah(\%)$  &$\ $ &$100$ &$\ $ &$75$ &$83.74$ \\
Input magnitude &$ac(\%)$  &$\ $ &$100$ &$\ $ &$71.25$ &$90.23$ \\
\hline
$nh=3$ &$ah(\%)$ &$\ $ &$80.56$ &$\ $ &$\ $ &$76.88$ \\
Input magnitude &$ac(\%)$ &$\ $ &$80.38$ &$\ $ &$\ $ &$81.25$ \\
\hline \hline
\end{tabular}
\caption{\label{tab_different_system}Average recovery rate and runtime in different networks}
\vspace{-5mm}
\end{table}

\begin{figure}
\centering
\includegraphics[width=3.5in]{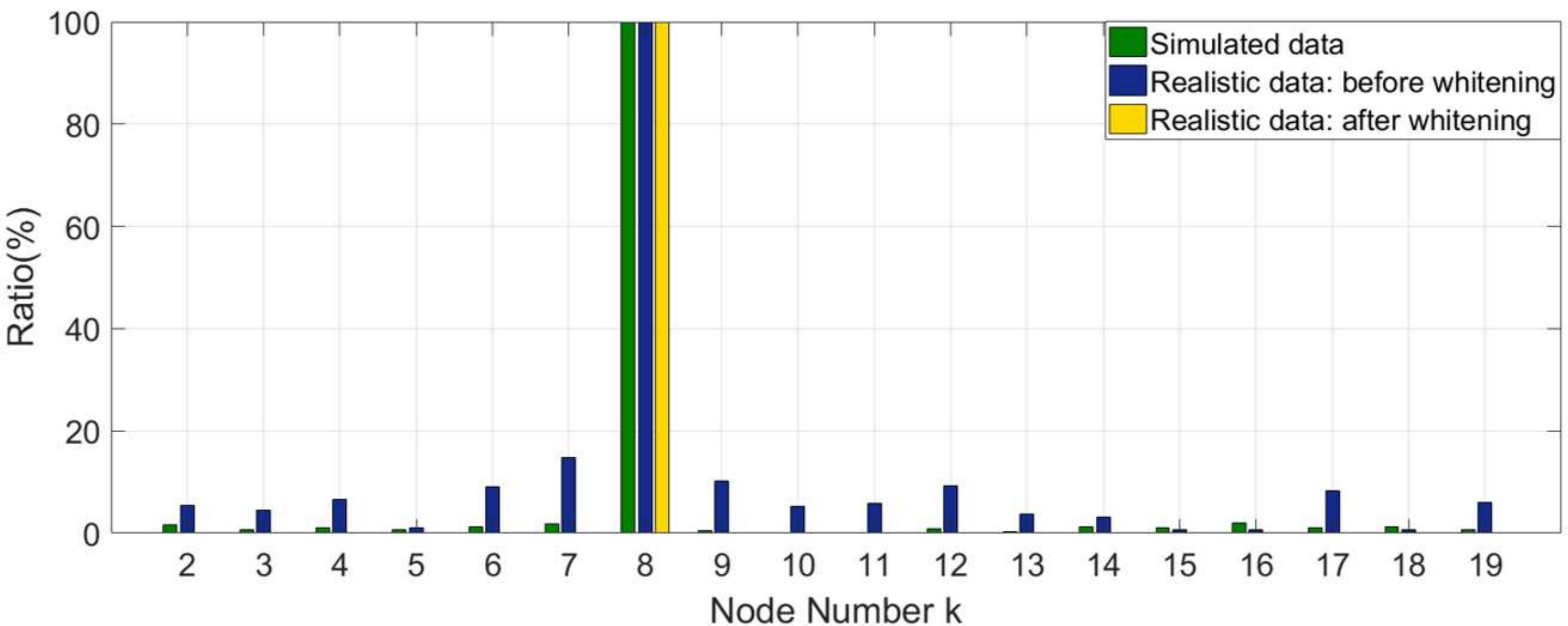}
\centering
\caption{This figure shows the ratio of $|{\Delta\mathbf{i}_8}^{H}\Delta\mathbf{i}_l|$ to $|{\Delta\mathbf{i}_8}^{H}\Delta\mathbf{i}_8|$. The simulated data and the Cholesky-whitened data are near uncorrelated while the realistic data before whitening is lowly correlated. Moreover, the sum of the ratio for non-whitened data (except node $8$) is $194.69\%$.}
\label{figure4-1-2}
\vspace{-6mm}
\end{figure}

% \begin{figure*}
% \subfigure[]{
% \begin{minipage}[t]{0.5\linewidth}
% \centering
% \includegraphics[width=3.5in]{Fig_distance_error_simdata}
% \caption{For the independent simulated data, the average errors are $1.83\%$ (real distance without whitening), $1.93\%$ (imaginary distance without whitening), $1.61\%$ (real distance with whitening) and $1.81\%$ (imaginary distance with whitening).}
% \end{minipage}}
% % \hspace{0.5in}
% \subfigure[]{
% \begin{minipage}[t]{0.5\linewidth}
% \centering
% \includegraphics[width=3.5in]{Fig_distance_error_realdata}
% \caption{For the real-world data, the average errors are $11.64\%$ (real distance without whitening), $9.45\%$ (imaginary distance without whitening), $3.35\%$ (real distance with whitening) and $3.31\%$ (imaginary distance with whitening).}
% \end{minipage}}
% \label{figure4-1-3}
% \end{figure*}

\begin{figure}
\centering
\subfigure[For the independent simulated data, the average errors are $1.83\%$ (real distance without whitening), $1.93\%$ (imaginary distance without whitening), $1.61\%$ (real distance with whitening) and $1.81\%$ (imaginary distance with whitening).]{
\includegraphics[width=3.5in]{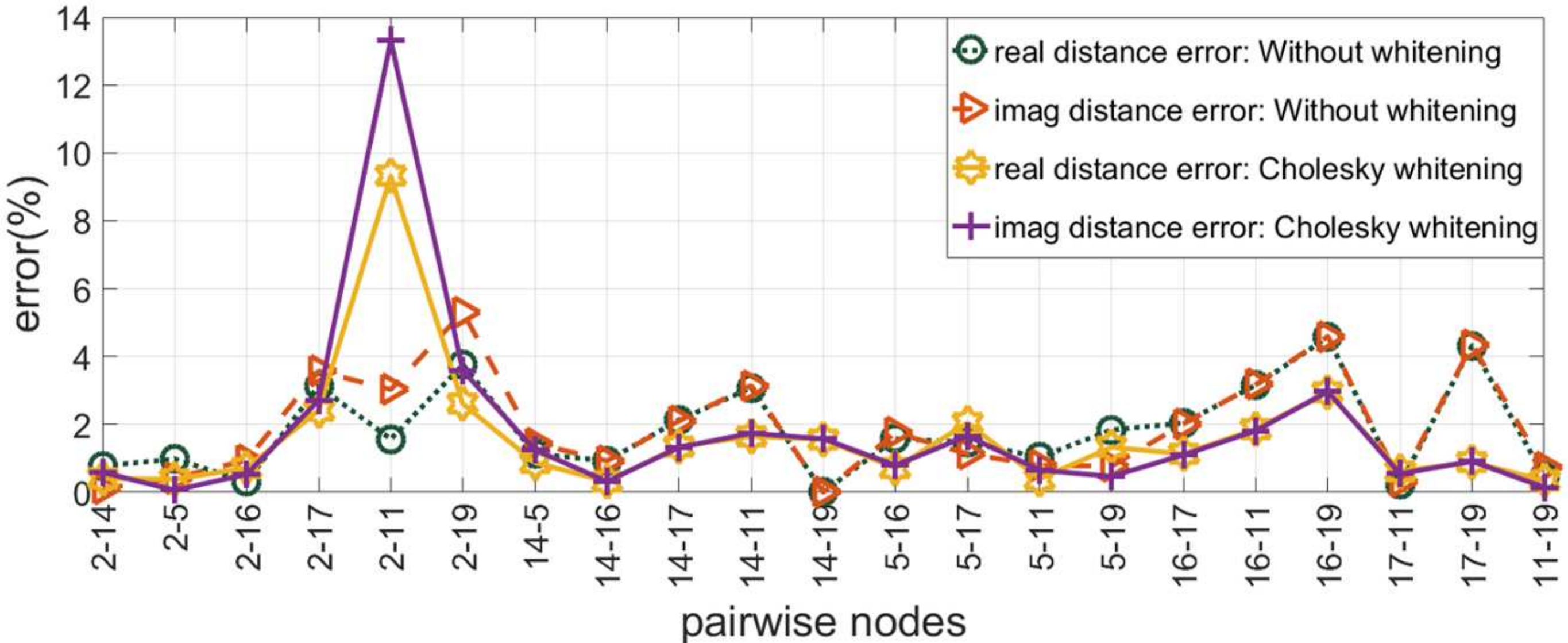}}
\hspace{0.5in}
\subfigure[For the real-world data, the average errors are $11.64\%$ (real distance without whitening), $9.45\%$ (imaginary distance without whitening), $3.35\%$ (real distance with whitening) and $3.31\%$ (imaginary distance with whitening).]{
\includegraphics[width=3.5in]{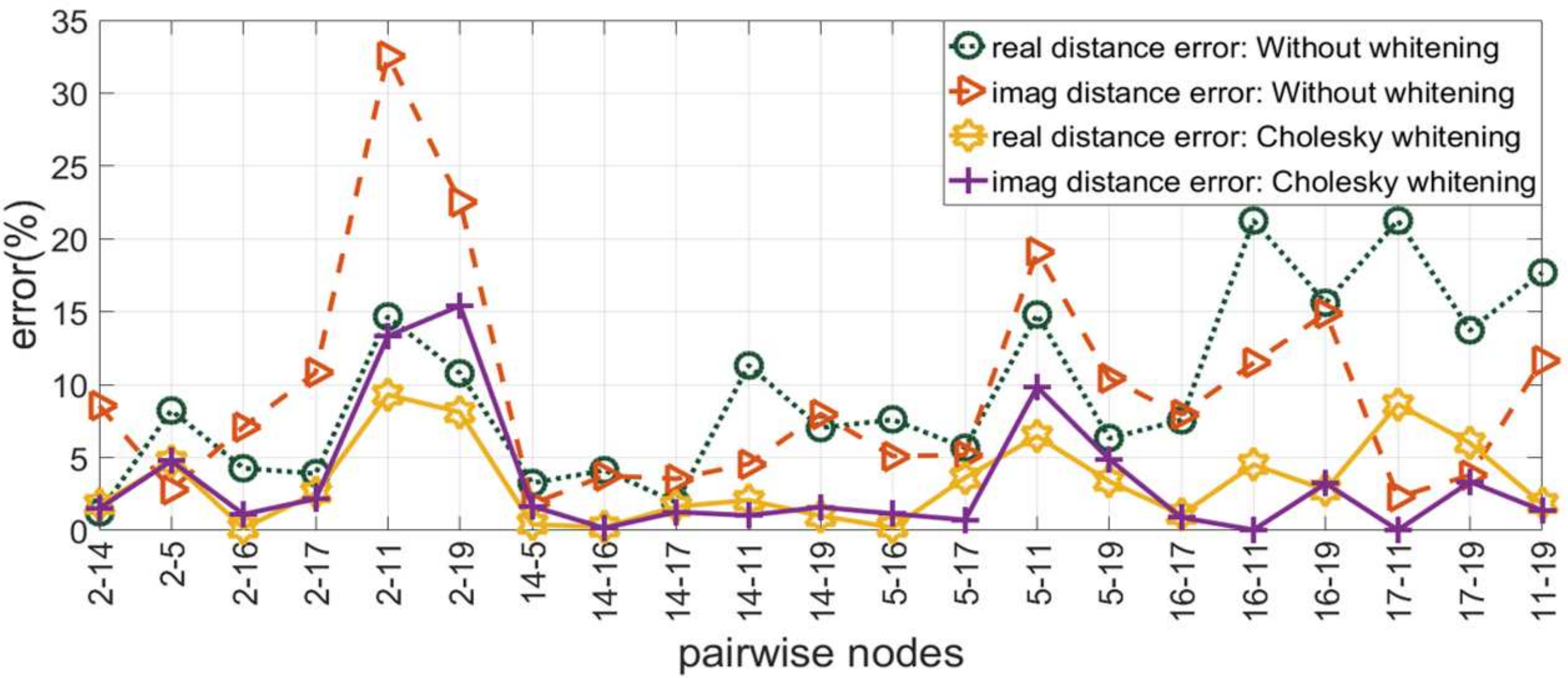}}
\centering
\caption{The distance-estimation errors of the observed nodes in $19$-bus system}
\label{figure4-1-3}
\end{figure}

\vspace{-4mm}
\subsection{Effectiveness of Whitening}
\label{Section4B}
In this section, we illustrate the effectiveness of whitening. In Fig. \ref{figure_19bus}, we consider $nh=5$, and the orange nodes are assumed to be known. The blue nodes are observed nodes and the white nodes are hidden nodes.

We first illustrate the error source of the distance estimation. Generally, the error comes from non-zero terms of inner product ${\Delta\mathbf{i}_b}^{H}\Delta\mathbf{i}_l,\ \forall b\neq l$. Fig \ref{figure4-1-2} demonstrates the ratio $|{\Delta\mathbf{i}_8}^{H}\Delta\mathbf{i}_l|/|{\Delta\mathbf{i}_8}^{H}\Delta\mathbf{i}_8|$. The result shows that the simulated data and the Cholesky-whitened data are near uncorrelated while the realistic data before whitening is slightly correlated. Moreover, the sum of the ratio for non-whitened data (except node $8$) is $194.69\%$, which shows low statistical correlations can lead to non-ignorable distance-estimation errors. However, Cholesky whitening process successfully enables the current-deviation measurements to be uncorrelated.

Fig. \ref{figure4-1-3} shows the absolute-percentage error of the estimated and true distance. We calculate them through the real and imaginary part, respectively. Generally, the distance estimation has high accuracy for simulated data in Fig. \ref{figure4-1-3} (a). As for real-world data (Fig. \ref{figure4-1-3} (b)), the imaginary distance after Cholesky whitening has the smallest average error (about $3.31\%$). Therefore, we utilize imaginary distance to continue the structure learning. Finally, as discussed in Table \ref{tab_19bus}, we find the correct hidden nodes and connections.
\vspace{-4mm}
\subsection{Performance Evaluation with Magnitude Data}
\label{Section4C}
\begin{figure}
\centering
\includegraphics[width=3.5in]{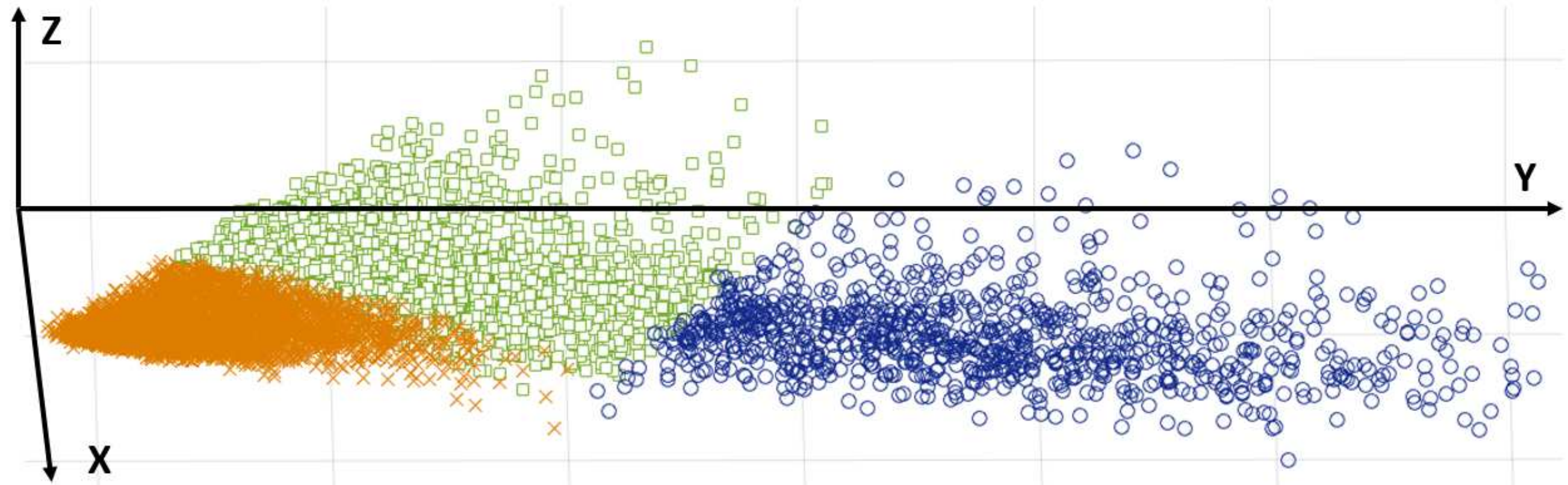}
\centering
\caption{$3D$ visualization for measurement points $z[n]$ in \ref{sec:data_clustering}. We find $3$ clusters (orange, green and blue clusters) for k-means when $k=3$.}
\label{figure_cluster}
\end{figure}

\begin{figure}
\centering
\includegraphics[width=3.5in]{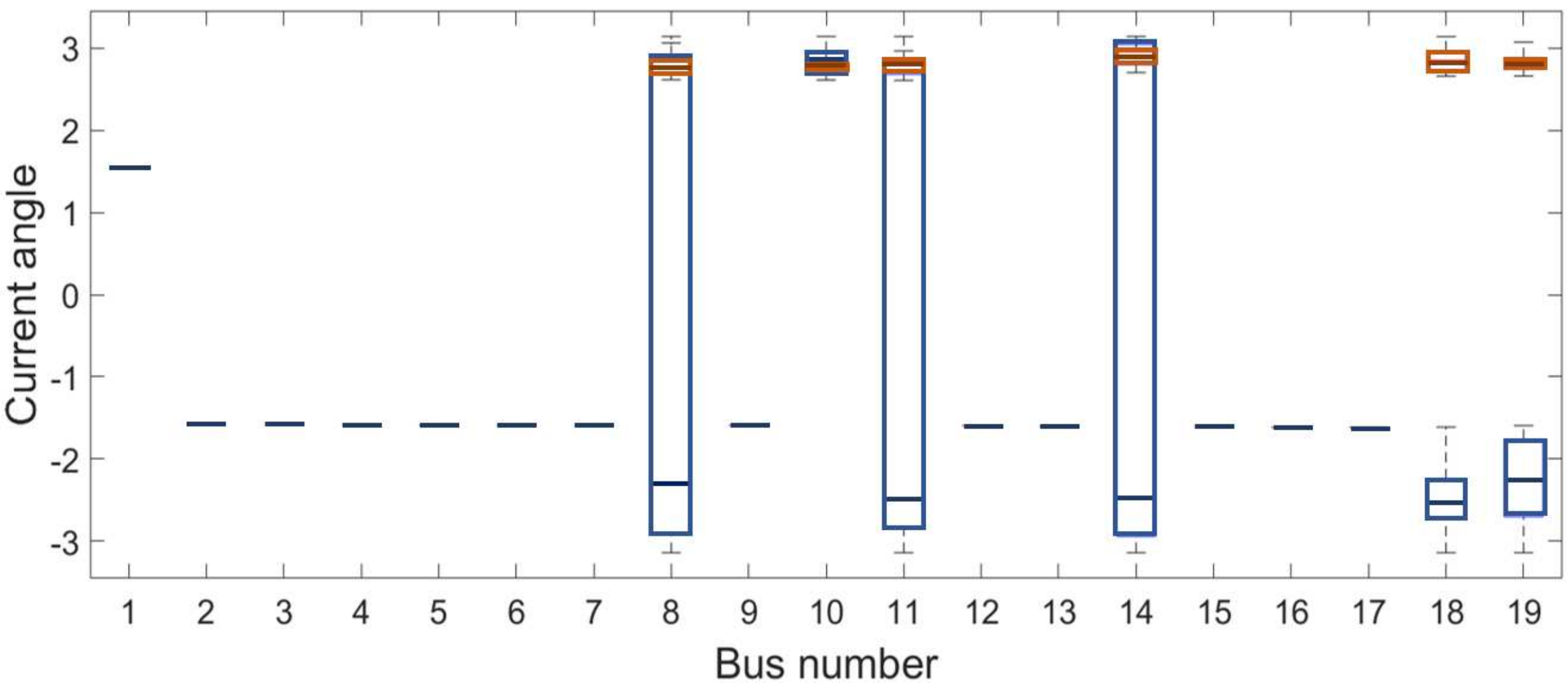}
\centering
\caption{The box diagram of current angles. We consider $19$ buses' current angles with $2$ collections of data: $1$) one year's data when $N=8760$, $2$) the orange clusters' data when $N=994$. The blue box represents the angle deviation calculated from one year's data and the orange box is obtained from the orange cluster.}
\label{figure_cluster_box}
\end{figure}
In this subsection, we illustrate that clustering to pick up angle-consistent data can give a good approximation for $|\mathbf{Z}_O|$ matrix in Section \ref{sec:data_clustering}.

Firstly, we show the clustering technique helps to eliminate angle information. With measurements of $19$-bus system, k-means clustering is employed when $k=3$ for the mapped points $z[n]$ in \ref{sec:data_clustering}. In this case, $z[n]$ is formed via the voltage magnitude and reactive power injections of observed buses in Fig. \ref{figure_19bus}. Fig. \ref{figure_cluster} visualizes the mapped data in $3D$ dimension with $3$ parts: orange, green and blue clusters.

To demonstrate the function of clustering, we draw the box diagram for current angles in the $19$-bus system. In Fig. \ref{figure_cluster_box}, we consider $2$ collections of data: $1$) one year's data when $N=8760$, $2$) the orange clusters' data when $N=994$. While the one year's data has large deviations in blue boxes, the orange cluster's data only has small changes in orange boxes.

\begin{figure}
\centering
\includegraphics[width=3.3in]{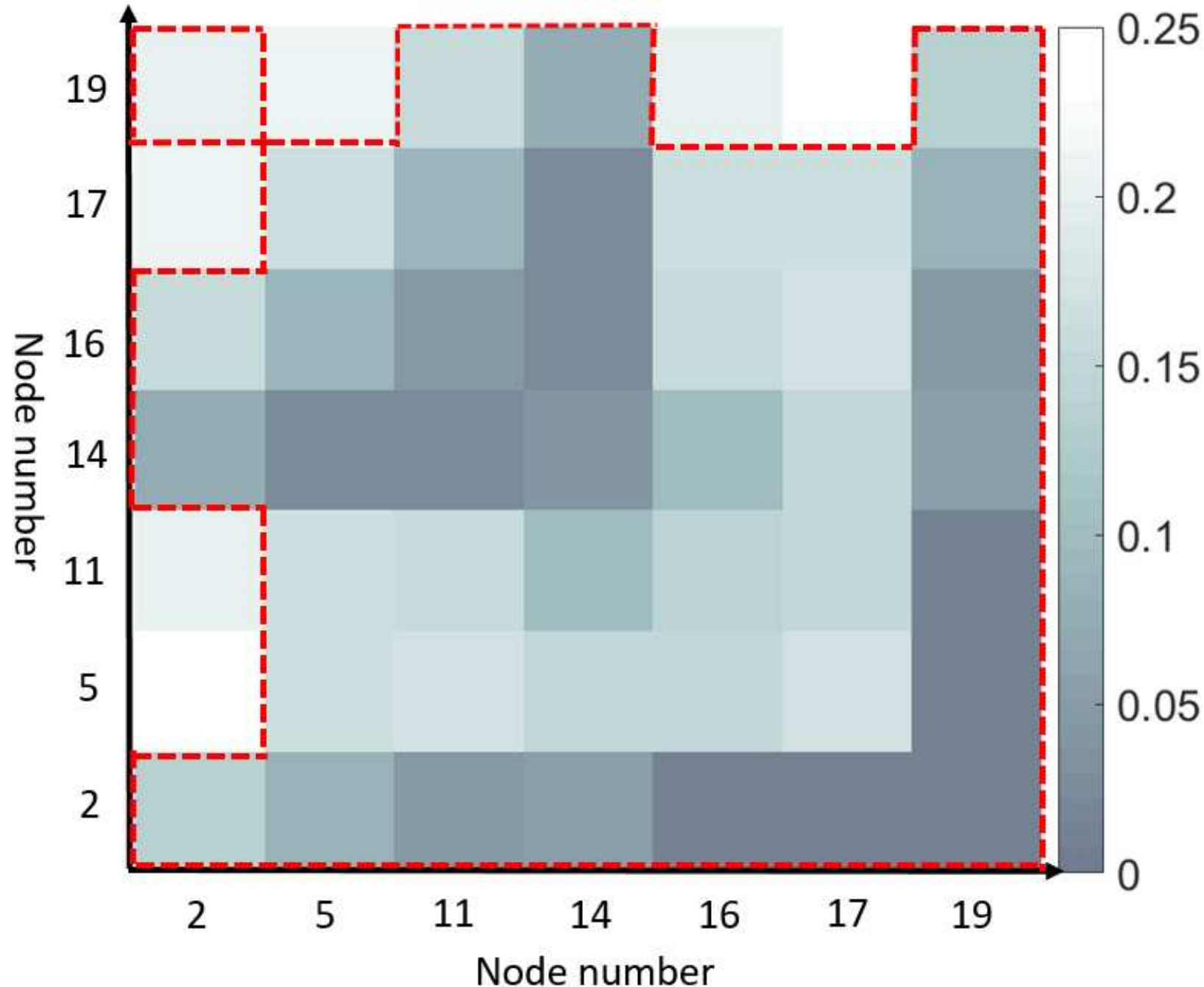}
\centering
\caption{The heat map of the error matrix for $|Z|_O$. Specifically, we calculate the element-wise percentage error between the estimated matrix in Section \ref{sec:data_clustering} and the true matrix. The maximum error is $23.48\%$ and most of the errors (in the dotted-red frame) are below $20\%$.}
\label{figure_heatmap}
\end{figure}

Subsequently, we illustrate the heat map of the error matrix for $|\mathbf{Z}|_O$, and the observed nodes are the same as Fig. \ref{figure_19bus}. Specifically, we calculate the element-wise percentage error between the estimated matrix in Section \ref{sec:data_clustering} and the true matrix in Fig. \ref{figure_heatmap}. The maximum error is $23.48\%$ and most of the errors (errors in the dotted-red frame) are below $20\%$.

Finally, we input voltage and current magnitudes of the orange cluster in Fig. \ref{figure_cluster} to the learning algorithm. The result is in Part \ref{Section4A}, Table \ref{tab_19bus} and \ref{tab_different_system}. The proposed algorithm still has a good performance for different hidden node combinations and various networks.

\vspace{-3mm}
\section*{Acknowledgement}
The first two authors would like to acknowledge the support and work done by Salt River Project.

\section{Conclusions}
\label{section5}
The distributed energy resources (DERs) are increasing in distribution grids, bringing high requirements for monitoring and control. To achieve these targets, the topology is the foundation for the system operator.

Due to frequent topology reconfiguration, this paper introduces a learning algorithm to identify the radial topology in real time. Starting from end nodes (observed nodes), the Recursive Grouping (RG) algorithm to detect hidden nodes among a target subset of nodes and recursively update the current target subset. In RG algorithm, line impedance is the metric to identify node relationships. To obtain the input impedances for RG (i.e., impedances among end nodes), we propose an estimation process. In this estimation, the correlation of measurements generates errors so we introduce Cholesky whitening to eliminate the correlation. Finally, we handle the cases when only voltage and current magnitudes are available or the network is three-phase unbalanced. We test our algorithms on various distribution systems with simulated data and real-world data and observe high performance in our proposed methods.

\bibliographystyle{IEEEtran}
\bibliography{IEEEabrv,myreport1}

\section{Appendix}
\label{section6}

\subsection{Line Impedance-based Structure Learning}
\label{appenA}
This subsection introduces Recursive Grouping (RG) algorithm that detects hidden nodes with partial nodes iteratively.

Due to the additivity of the distance, we introduce the following lemma \cite{ref:Choi2011l}:
\begin{lemma} \label{lemma2}
For distance $d_{ij}$, $\forall
i,j\in \mathcal{V}$ on a tree, the following two properties on $\Phi_{ijk}=d_{ik}-d_{jk}$ hold:
~\\
(i) $\Phi_{ijk}=d_{ij}$ for all $k\in \mathcal{V}\setminus\{i,j\}$ if and only if $i$ is a leaf node and $j$ is its parent.
~\\
(ii) $-d_{ij}<\Phi_{ijk}=\Phi_{ijk'}<d_{ij}$ for all $k,k'\in \mathcal{V}\setminus\{i,j\}$ if and only if both $i$ and $j$ are leaf nodes and they have the same parent, i.e., they belong to the same sibling group.
\end{lemma}

\begin{figure}
% * <liaoyizheng@gmail.com> 2019-01-02T06:39:29.464Z:
%
% > \begin{figure}
% > \centering
% > \includegraphics[width=3.5in]{Fig_RG_magnitude}
% > \centering
% > \vspace{-4mm}
% > \caption{Illustration of how to utilize estimated distance from voltage and current magnitudes to learn the structure.}
% > \label{figure_rg_magnitude}
% > \vspace{-4mm}
% > \end{figure}
%
% these two figures are not in 8-page right now
%
% ^.
\centering
\includegraphics[width=3.5in]{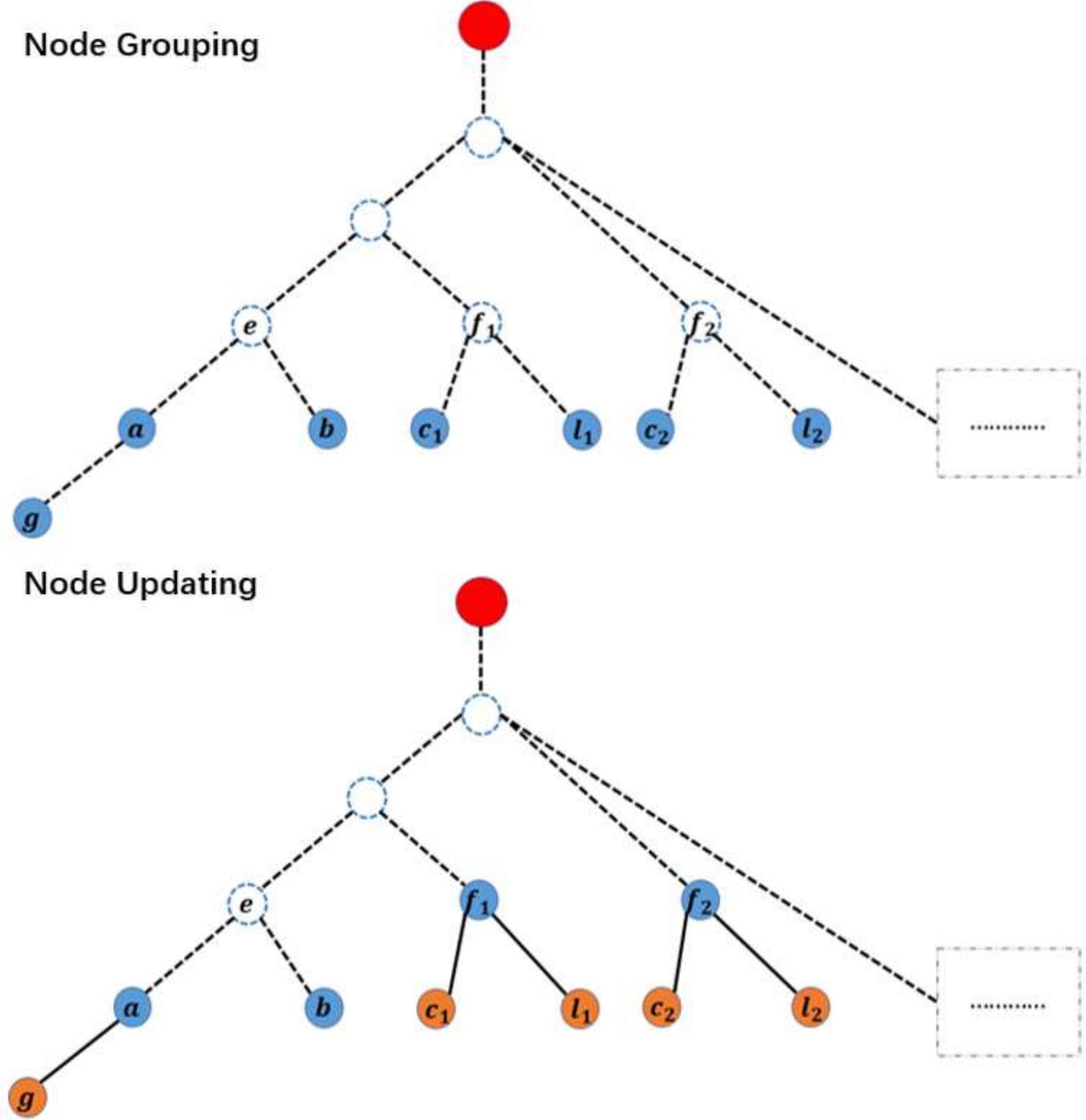}
\centering
\vspace{-4mm}
\caption{Illustration of RG algorithm.}
\label{figure_rg_magnitude}
\vspace{-4mm}
\end{figure}

Based on Lemma \ref{lemma2}, a subroutine called ``Node Grouping" can be implemented to group the current nodes and detect hidden nodes\cite{ref:Choi2011l}.
\begin{itemize}
\item If $\Phi_{ijk}=d_{ij}$ ($\forall k\in \mathcal{V}\backslash\{i,j\}$), $i$ is a leaf node and $j$ is a parent of $i$. Similarly, if $\Phi_{ijk}=-d_{ij}$ ($\forall k\in \mathcal{V}\backslash\{i,j\}$), $j$ is a leaf node and $i$ is a parent node of $j$.

\item If $\Phi_{ijk}(\forall k\in \mathcal{V}\backslash{i,j})$ is constant but not equal to either $d_{ij}$ or $-d_{ij}$, $i$ and $j$ are observed nodes and they are siblings.

% \item If $\Phi_{ijk}(\forall k\in \mathcal{V}\backslash{i,j})$ is not a constant. Node $i$ and $j$ may (a) not have siblings or parent-child relationship, (b) be siblings but at least one of them is not terminal, (c) have a child-parent relationship but the child is not terminal.
\end{itemize}

For example, we find in Fig. \ref{figure_rg_magnitude}, node $a$ is the parent of node $g$. Node $c_1$ and $l_1$ are siblings and share the same parent node $f_1$. The proof of Lemma \ref{lemma2} and ``Node Grouping" can be found in \cite{ref:Choi2011l}. ``Node Grouping" categorizes the current-node set $\mathcal{Y}$ into different partitions $\{\Pi_q\}_{q=1}^Q$. Any two nodes in an arbitrary $\Pi_q$ ($|\Pi_q|>2$) belong to one of the following types: (1) they are siblings and are observed nodes, (2) they have a parent-child relationship in which the child is observed. For some $q$, $\Pi_q$ may consist of a single node, like node $b$ in Fig. \ref{figure_rg_magnitude}. After this partition, we update the target set for further grouping process, i.e., ``Node Updating" in Fig. \ref{figure_rg_magnitude}.

The ``Node Updating" process is conducted via the following criteria: $1$) if the node in $\mathcal{Y}$ doesn't connect to any other node like node $b$ in Fig. \ref{figure_rg_magnitude}, we include it in the new target set $\mathcal{Y}_{new}$. $2$) If the node in $\mathcal{Y}$ is the parent node like node $a$ in Fig. \ref{figure_rg_magnitude}, we include it in $\mathcal{Y}_{new}$ since its relationship with the hidden nodes isn't figured out. $3$) The hidden nodes detected in the last ``Node Grouping" process are included in $\mathcal{Y}_{new}$ like node $f_1$ and $f_2$. Therefore, we construct the target node set in the next iteration, i.e., node $\{a,b,f_1,f_2\}$ in Fig. \ref{figure_rg_magnitude}.

To enable the updated nodes set to be ``observed", we need to recompute the distance between the hidden node $h\in \mathcal{Y}$ and all other nodes $p\in \mathcal{Y}$ in Step $7$. We denote $\mathcal{Y}_{old}$ to be the observed-node set in the previous iteration. Let $i,j\in \mathcal{C}(h)$ be two children of $h$, and let $k\in \mathcal{Y}_{old}\setminus\{i,j\}$ be an arbitrary node. Knowing that $d_{ih}-d_{jh}=d_{ik}-d_{jk}=\Phi_{ijk}$ and $d_{ih}+d_{jh}=d_{ij}$, we calculate the distance between $i$ and $h$ as follows:
\begin{equation}
 d_{ih}=\frac {1}{2}(d_{ij}+\Phi_{ijk}).\label{eqn:distcal1}
\end{equation}

For any other node $p\in \mathcal{Y}$, we compute $d_{hp}$ by discussing $p$ is hidden or not:
\begin{equation}\label{eqn:distcal2}
d_{hp}=
\left\{
\begin{aligned}
&d_{ip}-d_{ih},\ p\in \mathcal{Y}_{old}\\
&d_{ik}-d_{ih}-d_{pk},\ p\ is\ hidden,\ and\ k\in \mathcal{C}(p).\\
\end{aligned}
\right.
\end{equation}

\begin{figure*}
\centering
\centering
\subfigure[]{
\begin{minipage}{0.2\linewidth}
\includegraphics[width=1.2in]{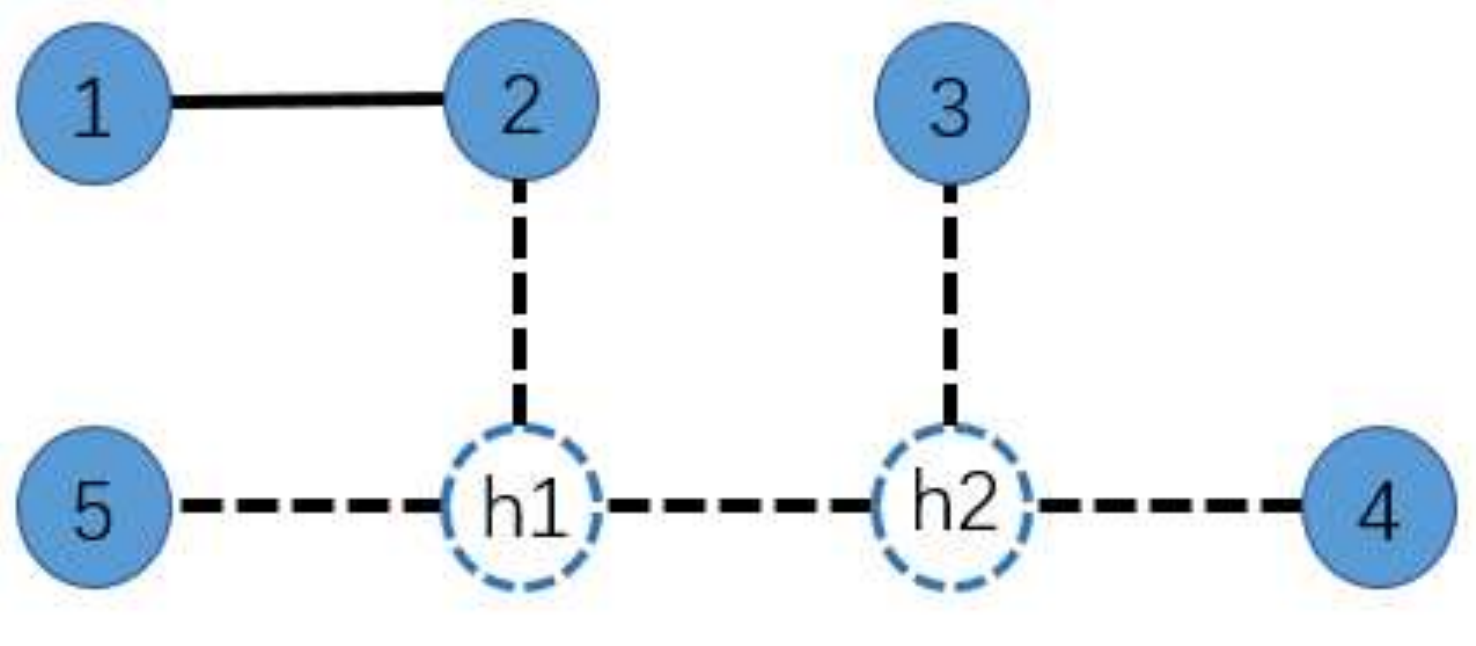}
\end{minipage}
}
\subfigure[]{
\begin{minipage}{0.2\linewidth}
\centering
\includegraphics[width=1.2in]{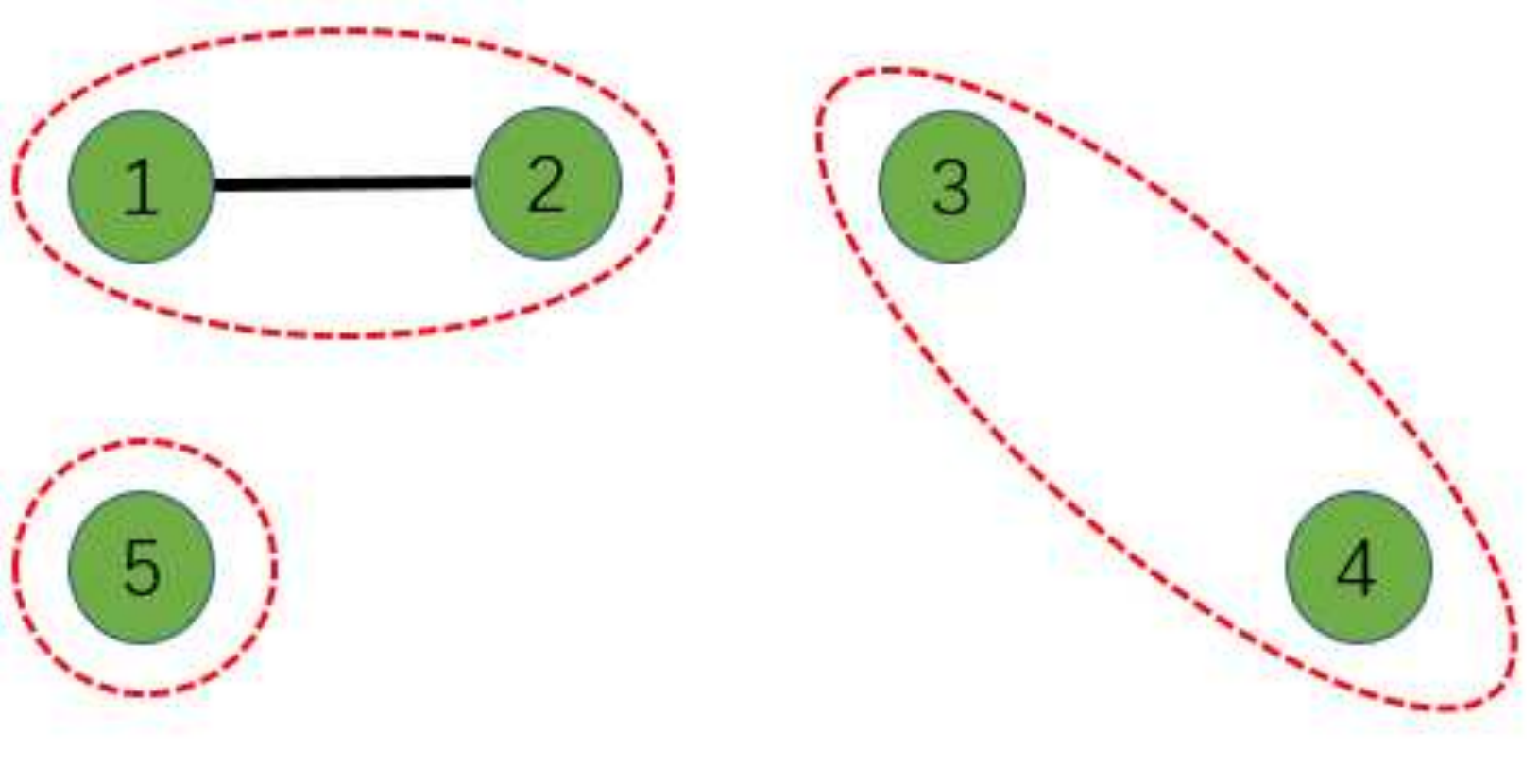}
\end{minipage}
}
\subfigure[]{
\begin{minipage}{0.2\linewidth}
\centering
\includegraphics[width=1.2in]{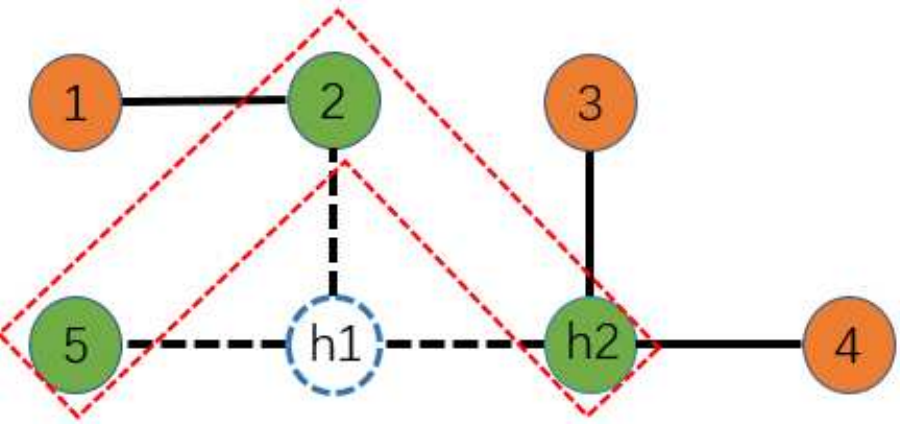}
\end{minipage}
}
\subfigure[]{
 \begin{minipage}{0.2\linewidth}
 \centering
 \includegraphics[width=1.2in]{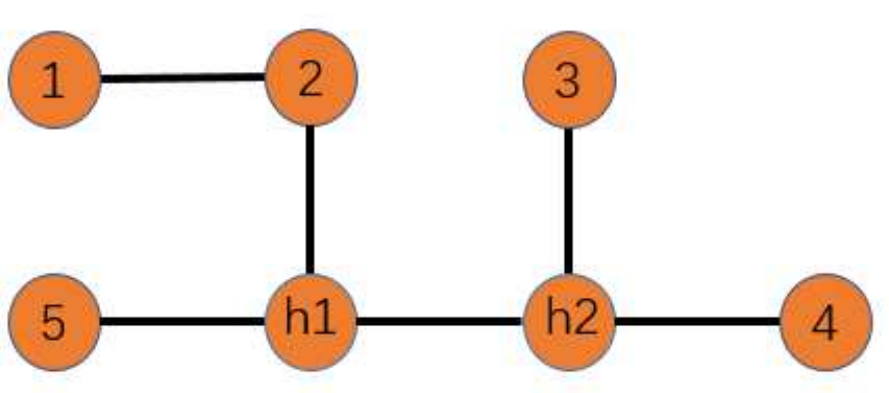}
 \end{minipage}
 }
\vspace{-3mm}
 \caption{Illustration of RG algorithm. (a) is the underlying topology to be identified. (b)-(d) are the identification process with RG. Green nodes represent the current-observed node set and orange nodes represent nodes that have been identified. Red circles mean the grouping of current-observed nodes via RG. After the first iteration in (b), we find a single node ($5$), a child-parent node pairs ($1$ and $2$) and a sibling group ($3$ and $4$). We detect a hidden node ($h2$) and update the current-observed node set in (c). In the next iteration, the current-observed nodes form a sibling group and we detect their parents ($h1$) in (d).}
 \label{FigRG}
 \vspace{-3mm}
\end{figure*}

Subsequently, we introduce the Recursive Grouping (RG) algorithm (Algorithm \ref{alg1}), termed as $\mathcal{RG}$. The iterative updating of $\mathcal{Y}$ in $\mathcal{RG}$ makes sure that relationships of all the nodes can be identified. For input distance matrix $\mathbf{D}$, $\mathbf{D}(a,b)$ and $\mathbf{D}(b,a)$ correspond to the distance $d_{ab}$. Fig. \ref{FigRG} illustrates the process of RG.

\begin{algorithm}
\caption{Recursive Grouping Algorithm for Structure Learning: $\mathcal{RG}$}
\label{alg1}
\begin{algorithmic}[1]
\REQUIRE Distances matrix D, Observed nodes set $\mathcal{O}$.
\ENSURE Structure of the latent tree and line parameter estimations.\\
\STATE Initialize $\mathcal{Y}:=\mathcal{O}$.
\WHILE {$|\mathcal{Y}|\geq3$}
\STATE Compute $\Phi_{ijk}=d_{ik}-d_{jk}$ for all $i,j,k\in \mathcal{Y}$.
\STATE Using the ``Node Grouping" procedure and partition $\mathcal{Y}$ into $\{\Pi_q\}_{q=1}^Q$, $\mathcal{Y}_{new}=\bigcup_{q:|\Pi_q|=1}\Pi_q$.
\STATE For each $q=1,\cdot \cdot \cdot,Q$ with $|\Pi_q|>2$, if $\Pi_q$ contains a parent node $u$, update $\mathcal{Y}_{new}=\mathcal{Y}_{new}\bigcup \{u\}$. Otherwise, introduce a new hidden node $h$, connect $h$ to every node in $\Pi_q$, and set $\mathcal{Y}_{new}=\mathcal{Y}_{new}\bigcup \{h\}$.
\STATE Update the observed nodes set: $\mathcal{Y}_{old}=\mathcal{Y}$ and $\mathcal{Y}=\mathcal{Y}_{new}$.
\STATE For each new hidden node $h\in \mathcal{Y}$, compute the distance $d_{hp}$ for all $p\in \mathcal{Y}$ using equation \eqref{eqn:distcal1} and \eqref{eqn:distcal2}.
\ENDWHILE
\IF{$|\mathcal{Y}|=2$}
\STATE Connect the two remaining nodes in $\mathcal{Y}$ with an edge and then stop.
\ELSIF {$|\mathcal{Y}|=1$}
\STATE Stop
\ENDIF
\end{algorithmic}
\end{algorithm}

\vspace{-2mm}

% \begin{remark}

% (1) The aforementioned algorithms are only suitable under the Assumption \ref{assum1}. If the intermediate nodes have a degree $2$, the ``Node Grouping" will not work. Our research domain focuses on identifying connectivities with $3$-degree intermediate nodes. {\color{red}In Section \ref{section3}, we will propose a network reduction technique to aggregate partial structures with $2$-degree intermediate nodes into one artificial node. This is beneficial to algorithm efficiency and distance-estimation accuracy.} (2) In terms of the sample data, the distance matrix $\widehat{D}$ is estimated by the voltage-current inner product. We can relax RG algorithm with tolerance in the implementation of `Node Grouping" and recalculation of distances. The relaxation is shown in \cite{ref:Choi2011l}.

% \end{remark}
\subsection{Proof of Theorem 1}
\label{appenB}
\begin{proof}
For the whitening process, we have:
\begin{equation} \label{appenA1}
\mathbf{M}\mathbf{M}^{H}=\mathbf{\Sigma},
\end{equation}
where $\mathbf{M}$ is the true Cholesky whitening matrix. \eqref{appenA1} is visualized as follows:
\begin{equation}
\label{appenA2}
\begin{aligned}
&\begin{bmatrix}
M_{11}&...&...&M_{1L}\\\ &...&...&...\\\ &M_{kk} &... &M_{kL}\\ \ &\ &... &... \\ \ &\ &\ &M_{LL}
\end{bmatrix}\cdot
\begin{bmatrix}
M_{11}&\ &\ \\... &... &\ &\ \\
M_{k1}^{*} &... &M_{kk} &\ \\... &... &... &\ \\M_{1L}^{*}  &... &M_{kL}^{*} &M_{LL}
\end{bmatrix}
\\=&
\begin{bmatrix}
\Sigma_{11}&...&...&...&\Sigma_{1L}\\\Sigma_{21}&...&...&...&\Sigma_{2L}\\... &... &\Sigma_{kk}&...&\Sigma_{kL}\\
\\\Sigma_{L1} &... &\Sigma_{kL}&...&\Sigma_{LL}
\end{bmatrix}_{\textstyle .}
\end{aligned}
\end{equation}

We arrange the bus number of the observed nodes set $\mathcal{O}$ from $k$ to $L$. The observed covariance matrix, termed as $\widehat{\mathbf{\Sigma}}_O$, contains value $\Sigma_{kk},\Sigma_{k,k+1},\cdot \cdot \cdot,\Sigma_{LL}$ that can be directly calculated with the observed measurements. Therefore, $\widehat{\mathbf{\Sigma}}_O=\mathbf{\Sigma}_O$. $\widehat{\mathbf{\Sigma}}_O$ is also conjugate-symmetric and uniquely Cholesky-decomposable. If we assume Cholesky decomposition of $\widehat{\mathbf{\Sigma}}_O$ is $\widehat{\mathbf{\Sigma}}_O=\widehat{\mathbf{M}}_O\widehat{\mathbf{M}}_O^H$ ($\widehat{\mathbf{M}}_O$ is a unique upper-diagonal matrix), we can conclude that:
\begin{equation*}
\widehat{\mathbf{M}}_O=
\begin{aligned}
\begin{bmatrix}
M_{kk} &... &M_{kL}\\ \ &... &...\\ \ &\ &M_{LL}
\end{bmatrix}_{\textstyle .}
\end{aligned}
\end{equation*}

Then, $\widehat{\mathbf{M}}_O=\mathbf{M}_O$. Given that $\mathbf{W}=\mathbf{M}^{-1}$ is also an upper-diagonal matrix, we rewrite $\mathbf{M}\mathbf{W}=\mathbf{Id}$ ($\mathbf{Id}$ is the identity matrix) in blocks form:
\begin{equation*}
\begin{aligned}
\begin{bmatrix}
\mathbf{M}_1 & \mathbf{M}_2\\ \ &\widehat{\mathbf{M}}_O
\end{bmatrix} \cdot
\begin{bmatrix}
\mathbf{W}_1 &\mathbf{W}_2\\ \ &\mathbf{W}_O
\end{bmatrix}
=
\begin{bmatrix}
\mathbf{Id}_1 &\ \\ \ &\mathbf{Id}_O
\end{bmatrix}_{\textstyle .}
\end{aligned}
\end{equation*}

Therefore, we have $\mathbf{W}_O={\widehat{\mathbf{M}}_O}^{-1}=\widehat{\mathbf{W}}_O$. This conclusion helps us to rewrite the whitening process:
\begin{equation*}
\begin{aligned}
\begin{bmatrix}
(\tilde{\mathbf{I}}_1)^T \\ (\tilde{\mathbf{I}}_O)^T
\end{bmatrix} =
\begin{bmatrix}
\mathbf{W}_1 &\mathbf{W}_2\\ \ &\widehat{\mathbf{W}}_O
\end{bmatrix}
\cdot
\begin{bmatrix}
(\mathbf{I}_1)^T \\ (\mathbf{I}_O)^T
\end{bmatrix}_{\textstyle ,}
\end{aligned}
\end{equation*}
where $(\tilde{\mathbf{I}}_1)^T$ and $(\mathbf{I}_1)^T$ correspond to measurements of unobserved nodes and $(\tilde{\mathbf{I}}_O)^T$ and $(\mathbf{I}_O)^T$ correspond to measurements of observed nodes. Considering \eqref{eqn:obserwhite}, we have: $(\widehat{\tilde{\mathbf{I}}}_O)^T=\widehat{\mathbf{W}}_O(\mathbf{I}_O)^T=(\tilde{\mathbf{I}}_O)^T$.
\end{proof}

\subsection{Proof of Theorem 2}
\label{appenC}
\begin{proof}
We can rewrite \eqref{appenA2} in a block-matrix form:
\begin{equation} \label{eqn19}
\begin{aligned}
\begin{bmatrix}
\mathbf{M}_1 &\mathbf{M}_2\\ \ &\widehat{\mathbf{M}}_O
\end{bmatrix} \cdot
\begin{bmatrix}
\mathbf{M}_1^{H} &\ \\ \mathbf{M}_2^{H} &\widehat{\mathbf{M}}_O^{H}
\end{bmatrix}
=
\begin{bmatrix}
\mathbf{\Sigma}_1 &\mathbf{\Sigma}_2 \\ \mathbf{\Sigma}_3 &\mathbf{\Sigma}_4
\end{bmatrix}_{\textstyle ,}
\end{aligned}
\end{equation}
where $\mathbf{\Sigma}_4$ is the covariance matrix for current deviations of observed nodes, and $\mathbf{\Sigma}_2$ is the covariance matrix of current deviations between observed nodes and hidden nodes. If current-deviation correlations only exist among observed nodes (Assumption \ref{assum3}), we can assume $\mathbf{\Sigma}_4$ is a dense matrix and $\mathbf{\Sigma}_2=\mathbf{0}_{(k-1)\times (L-k+1)}$, $\mathbf{\Sigma}_3=\mathbf{0}_{(L-k+1)\times (k-1)}$.

According to \eqref{eqn19}, we have:

\begin{equation}\label{eqn:block_matrix_derivation}
\left\{
\begin{aligned}
&\widehat{\mathbf{M}}_O\times \mathbf{M}_2^H=\mathbf{0}_{(L-k+1)\times (k-1)},\\
&\mathbf{M}_2\times\widehat{\mathbf{M}}_O^{H}=\mathbf{0}_{(k-1)\times (L-k+1)}.\\
\end{aligned}
\right.
\end{equation}

Then we have: $\mathbf{M}_2\times\widehat{\mathbf{M}}_O^{H} \times\widehat{\mathbf{M}}_O\times \mathbf{M}_2^H =\mathbf{0}_{(k-1)\times(k-1)}$, which implies:
\begin{equation*}
\begin{aligned}
\mathbf{M}_2\times\mathbf{\Sigma}_4^H\times \mathbf{M}_2^H=\mathbf{0}_{(k-1)\times(k-1)},\\
\mathbf{M}_2 \times \mathbf{V}\times \mathbf{\Lambda} \times \mathbf{V}^H \times \mathbf{M}_2^H = \mathbf{0}_{(k-1)\times(k-1)},\\
\mathbf{M}_2 \times \mathbf{V}\times \mathbf{\Lambda} \times (\mathbf{M}_2\times  \mathbf{V})^H = \mathbf{0}_{(k-1)\times(k-1)},
\end{aligned}
\end{equation*}
where $\mathbf{V}$ is a $(L-k+1)\times(L-k+1)$ matrix containing all the eigenvectors of $\mathbf{\Sigma}_4$ and $\mathbf{\Lambda}$ is a diagonal matrix with positive eigenvalues.
Since $\mathbf{M}_2\times \mathbf{V}$ and $(\mathbf{M}_2\times \mathbf{V})^H$ are conjugate symmetric and $\mathbf{\Lambda}$ is a diagonal matrix with positive values, we know:
\begin{equation} \label{eqn:m2v}
\mathbf{M}_2\times \mathbf{V} = \mathbf{0}_{(k-1)\times(L-k+1)}.
\end{equation}

According to the orthogonal property of eigenvectors, we have: $\mathbf{V}\times \mathbf{V}^H=\mathbf{Id}_{(L-k+1)\times(L-k+1)}$ ($\mathbf{Id}$ is the identity matrix). We use $\mathbf{V}^H$ right multiply \eqref{eqn:m2v}, and finally, we can obtain $\mathbf{M}_2 = \mathbf{0}_{(k-1)\times (L-k+1)}$.

\end{proof}

\subsection{Proof of Feasibility for Structure Learning without Angle}
\label{appenD}
\begin{proof}
 Fig. \ref{figure_rg_magnitude} illustrates the learning process. Firstly, we consider observed nodes set $\mathcal{O}$ as the current-active set $\mathcal{Y}=\mathcal{O}$. We include nodes $a, b, c_1, l_1, c_2, l_2, g\in \mathcal{O}$ to represent all the possibilities of observed nodes.

According to Lemma \ref{lemma2} in Appendix \ref{appenA}, we first verify the ability to identify the parent-child relationship of node $a$ and $g$, namely, the correctness of Lemma \ref{lemma2} $(i)$. For every $k\in \mathcal{O}\setminus \{a,g\}$, $\Phi_{agk}=d_{ak}-d_{gk}=|\mathbf{Z}(a,a)|-|\mathbf{Z}(g,g)|$,
where the last equality holds by $\mathbf{Z}(a,k)=\mathbf{Z}(g,k)$ from Lemma \ref{lemma1}. Similarly, the distance between nodes $a$ and $g$ is $d_{ag}=|\mathbf{Z}(g,g)|-|\mathbf{Z}(a,a)|$
by $\mathbf{Z}(a,a)=\mathbf{Z}(a,g)$. Thus, for arbitrary $k\in \mathcal{O}\setminus \{a,g\}$, we have $\Phi_{agk}=d_{ag}=-\Phi_{gak}$, which holds if $a$ is the parent of $g$. The above proof is applicable to every parent-child node pair when the child node is a leaf node.

For the other direction of Lemma \ref{lemma2} $(i)$, we prove the contraposition. There exists non-parent-child node pair $g$ and $b$, and another observed node $c_1$ such that $\Phi_{bgc_1}-d_{bg}=d_{bc_1}-d_{gc_1}-d_{bg}=2(|\mathbf{Z}(b,g)|-|\mathbf{Z}(g,g)|)<0$ by $\mathbf{Z}(b,c_1)=\mathbf{Z}(g,c_1)$ and $|\mathbf{Z}(b,g)|<|\mathbf{Z}(g,g)|$. In general, Lemma \ref{lemma2} $(i)$ holds for our defined distance.

Then, we verify Lemma \ref{lemma2} $(ii)$ for leaf nodes $c_1$ and $l_1$ in the sibling group. For every $k\in \mathcal{O}\setminus \{c_1,l_1\}$, we have $\Phi_{c_1l_1k}=d_{c_1k}-d_{l_1k}=|\mathbf{Z}(c_1,c_1)|-|\mathbf{Z}(l_1,l_1)|$,
where the last equality holds by $\mathbf{Z}(c_1,k)=\mathbf{Z}(l_1,k)$. Thus, no matter which $k$ we choose, $\Phi_{c_1l_1k}$ is a constant. Similarly, the distance between $c_1$ and $l_1$ is:
\begin{equation*}
\begin{aligned}
d_{c_1l_1}&=|\mathbf{Z}(c_1,c_1)|+|\mathbf{Z}(l_1,l_1)|-2|\mathbf{Z}(c_1,l_1)|,\\
      &\geq |(|\mathbf{Z}(c_1,c_1)|-|\mathbf{Z}(c_1,l_1)|)-(|\mathbf{Z}(l_1,l_1)|-|\mathbf{Z}(c_1,l_1)|)|,\\
      &\geq |\Phi_{c_1l_1k}|.
\end{aligned}
\end{equation*}

The above proof is applicable for every $2$ leaf nodes when they are in the same sibling group.

Then, we prove the other direction of Lemma \ref{lemma2} $(ii)$. Consider the contraposition, there exist non-sibling node pair $b$ and $c_1$, and another observed node $k\in \mathcal{O}\setminus \{b,c_1\}$ such that:
$\Phi_{bc_1k}=|\mathbf{Z}(c_1,c_1)|+2(|\mathbf{Z}(c_1,k)|-|\mathbf{Z}(b,k)|)$. Therefore, if $k$ is varying, $\Phi_{bc_1k}$ is not a constant. In general, Lemma \ref{lemma2} $(ii)$ holds for our defined distance.

Secondly, using Lemma \ref{lemma2}, we find the parent-child node pair $a$ and $g$, sibling group $c_1$ and $l_1$ with a hidden parent $f_1$, and a single node $b$. Without loss of generality, they represent all the possible groups for $\mathcal{O}$ by RG algorithm in Appendix \ref{appenA}.

Subsequently, nodes $a$, $b$, and $f_1$ represent all possible types to form the new current-active set $\mathcal{Y}_{new}$. To utilize the induction idea to prove the feasibility of our defined distance in RG algorithm, we only need to prove distances among the new set $\mathcal{Y}_{new}$ still have the same form defined as $d_{hk}=|\mathbf{Z}(h,h)|+|\mathbf{Z}(k,k)|-2|\mathbf{Z}(h,k)|,\ \forall h,k\in \mathcal{Y}_{new}$. However, there are $3$ types of distance: $1$) the distance between two nodes in $\mathcal{Y}$, $2$) the distance between a node in $\mathcal{Y}$ and a detected hidden node, and $3$) the distance between $2$ detected hidden node. Thus, we include another sibling group $c_2$, $l_2$ and their parent $f_2$ to form the type $3$) distance $d_{f_1f_2}$.

For distance type $1$) like $d_{ab}$, it is directly calculated by the distance definition. For distance type $2$), without loss of generality, we consider $d_{k_1f_1},\ \forall k_1\in \mathcal{Y}\cap \mathcal{Y}_{new}$, defined as $d_{k_1f_1}=d_{k_1c_1}-d_{c_1f_1}$. Since $\Phi_{c_1l_1k},\ \forall k\in \mathcal{Y}$ is a constant, we employ definition in \eqref{eqn:distcal1}, Appendix \ref{appenA} to calculate $d_{c_1f_1}$: $d_{c_1f_1}=\frac{1}{2}(\Phi_{c_1l_1k}+d_{c_1l_1})=|\mathbf{Z}(c_1,c_1)|-|\mathbf{Z}(c_1,f_1)|$, where the last equality holds by $\mathbf{Z}(c_1,l_1)=\mathbf{Z}(c_1,f_1)$. Therefore, we calculate $d_{k_1f_1},\ \forall k_1\in \mathcal{Y}\cap \mathcal{Y}_{new}$:
\vspace{-3mm}
\begin{equation*}
\begin{aligned}
d_{k_1f_1}&=d_{k_1c_1}-d_{c_1f_1},\\          &=|\mathbf{Z}(k_1,k_1)|+|\mathbf{Z}(c_1,f_1)|-2|\mathbf{Z}(k_1,c_1)|,\\ &=|\mathbf{Z}(k_1,k_1)|+|\mathbf{Z}(f_1,f_1)|-2|\mathbf{Z}(k_1,f_1)|,
\end{aligned}
\end{equation*}

where the last equality holds by $\mathbf{Z}(c_1,f_1)=\mathbf{Z}(f_1,f_1)$ and $\mathbf{Z}(k_1,c_1)=\mathbf{Z}(k_1,f_1)$.

Furthermore, we consider $d_{f_1f_2}$ for distance type $3$).
\vspace{-3mm}
\begin{equation*}
\begin{aligned}
d_{f_1f_2}&=d_{c_1c_2}-d_{c_1f_1}-d_{c_2f_2},\\         &=|\mathbf{Z}(c_1,f_1)|+|\mathbf{Z}(c_2,f_2)|-2|\mathbf{Z}(c_1,c_2)|,\\         &=|\mathbf{Z}(f_1,f_1)|+|\mathbf{Z}(f_2,f_2)|-2|\mathbf{Z}(f_1,f_2)|,
\end{aligned}
\end{equation*}

where the last equality holds by $\mathbf{Z}(c_1,f_1)=\mathbf{Z}(f_1,f_1)$, $\mathbf{Z}(c_2,f_2)=\mathbf{Z}(f_2,f_2)$ and $\mathbf{Z}(c_1,c_2)=\mathbf{Z}(f_1,f_2)$. In general, we prove the $3$ types of distances have the same form as defined. Therefore, Lemma \ref{lemma2} and the node grouping criterion hold on the current-active set $\mathcal{Y}_{new}$. Thus, we prove node updating with defined distance is correct and by induction method, our defined distance helps the RG algorithm to recover the tree.
\end{proof}

\ifCLASSOPTIONcaptionsoff
  \newpage
\fi

%\appendices
%\section{The algorithm of}
%\label{sec:hidval}
%abc
%\subsection{aaa}
%aaa

% End of Appendix

\vspace{-10 mm}
\end{document}